\documentclass[twocolumn,superscriptaddress,preprintnumbers,aps,prb,citeautoscript]{revtex4-2}

\usepackage[colorlinks]{hyperref}
\usepackage{amsmath}
\usepackage{graphicx}
\usepackage{latexsym}
\usepackage{amsmath,amssymb}
\usepackage{bm} 
\usepackage{xcolor}
\usepackage{stackrel}
\usepackage{accents}
\usepackage{braket}
\usepackage[normalem]{ulem}
\usepackage[utf8]{inputenc}
\usepackage{notes2bib}

\bibnotesetup{
note-name = ,
use-sort-key = false
}

\newcommand{\rd}{\mathrm{d}}
\newcommand{\ri}{\mathrm{i}}
\newcommand{\re}{\mathrm{e}}

\newcommand{\lrangle}[1]{\left< #1\right>} 

\begin{document}
\title{Topological Quantum Computation on a Chiral Kondo Chain}
\author{Tianhao Ren}
\email{tren@bnl.gov}
\affiliation{Condensed Matter Physics and Materials Science Division, Brookhaven National Laboratory, Upton, New York 11973, USA}
\author{Elio J. K\"onig}
\affiliation{Max-Planck Institute for Solid State Research, 70569 Stuttgart, Germany}
\author{Alexei M. Tsvelik}
\email{atsvelik@bnl.gov}
\affiliation{Condensed Matter Physics and Materials Science Division, Brookhaven National Laboratory, Upton, New York 11973, USA}

\begin{abstract}
    We describe  the chiral Kondo chain model based on the symplectic Kondo effect and demonstrate that it has a quantum critical ground state populated by  non-Abelian anyons. We show that the fusion channel of two arbitrary anyons can be detected by locally coupling the two anyons to an extra single channel of chiral current and measuring the corresponding conductance at finite frequency. Based on such measurements, we propose that the chiral Kondo chain model with symplectic symmetry can be used for implementation of measurement-only topological quantum computations, and it possesses a number of distinct features favorable for such applications. The sources and effects of errors in the proposed system are analyzed, and possible material realizations are discussed.
\end{abstract}

\maketitle

\section{Introduction}
 
The design of quantum devices and harvesting their properties for quantum technological utilization is a prime application of present day condensed matter research. In this context, the topological quantum computation \cite{FaultTolerant,NayakDasSarma,PachosBook2012,10.21468/SciPostPhys.3.3.021} is particularly appealing, because it allows for intrinsically robust storage and processing of information by braiding of anyons. The latter occur  are low-energy quasiparticles occuring  in strongly interacting many-body systems which display generic exchange statistics and  an irrational Hilbert space dimension (known as ``quantum dimension''). 
 
Existing proposals for platforms to realize topological quantum computation include fractional quantum Hall states using non-Abelian anyonic excitations \cite{NayakDasSarma} and one-dimensional wire networks using Majorana zero modes \cite{Kitaev_2001,DasSarma}. Besides, solitary anyons bound to the impurity site are predicted to exist in the ground states of various quantum critical Kondo models and numerous experimental observations of quantum criticality \cite{PotokGoldhaberGordon2007,KellerGoldhaberGordon2015,Iftikhar2015,Iftikhar2018,PouseGoldhaberGordon2021,PhysRevLett.130.146201} indirectly support this claim. Topological quantum computation, however, require multiple anyons. It was suggested by Sela \textit{et al.} \cite{Sela1,Sela2,Sela3} that multiple anyons may exist in a chiral multichannel Kondo chain where multiple impurities are placed on the chiral edge of a topologically nontrivial bulk. In the absence of backscattering, the chiral electrons do not generate the Ruderman–Kittel–Kasuya–Yosida (RKKY) interaction between the impurities, and the critical ground states propelled by the Kondo screening develop without hindrance. The resulting Hilbert space of the ground states is spanned by fusion of entangled anyons located at the impurity sites\cite{Komijani2020}. The braiding of anyons can be performed without their physical permutation by means of measurement-only topological quantum computation \cite{BondersonNayak2008}.

 The existing schemes for the construction of quantum gates, such as the ones described in Refs. \onlinecite{Sela1,Sela2,Sela3}, are based on the multichannel Kondo models. The corresponding quantum critical state is unstable with respect to channel anisotropy and hence  requires fine tuning of the energy levels and the exchange interactions of different channels. Although the experimental realizations of two- and three-channel Kondo criticality \cite{PotokGoldhaberGordon2007,KellerGoldhaberGordon2015,Iftikhar2015,Iftikhar2018,PouseGoldhaberGordon2021,PhysRevLett.130.146201} demonstrate that this problem can be ameliorated, it is still desirable to consider alternate realizations of Kondo quantum critical point (QCP).

In this paper we describe a chiral Kondo chain model with symplectic symmetry possessing non-Abelian anyons in its ground state. We also describe a measurement-only protocol for universal quantum computations. This model may be easier to implement, although it also requires fine tuning since, contrary to our original beliefs \cite{PhysRevB.107.L201401,KoenigTsvelik2023}, there are perturbations which drive the model to a Fermi liquid. As for the implementation, we envisage that the role of quantum impurities will be played by mesoscopic devices consisting of a mesoscopic superconducting Coulomb box in proximity to several quantum dots.

The paper is organized as follows. In Sec. \ref{sec:model} we review the symplectic Kondo effect, which is central to our construction of the chiral Kondo chain model. In Sec. \ref{sec:integrability} we define the chiral Kondo chain model with symplectic symmetry and show its integrability. In Sec. \ref{sec:lowenergy} we describe the low energy properties of the chiral Kondo chain model and show that it possesses localized non-Abelian anyons in its ground state in the sense of local operators. In Sec. \ref{sec:correlation} we discuss how the information encoded by the anyons can be extracted by measuring the correlation functions. In Sec. \ref{sec:TQC} we describe the implementation of the chiral Kondo chain with symplectic symmetry as a platform for universal quantum computations. In Sec. \ref{sec:summary} we provide a summary of our results. Several basic ingredients of anyonic quantum computation are reviewed in the appendices to make the paper self-contained.

\section{Review of the Symplectic Kondo Effect}
\label{sec:model}
 
\begin{figure}[htp!]
    \centering
    \includegraphics[width=0.5\linewidth]{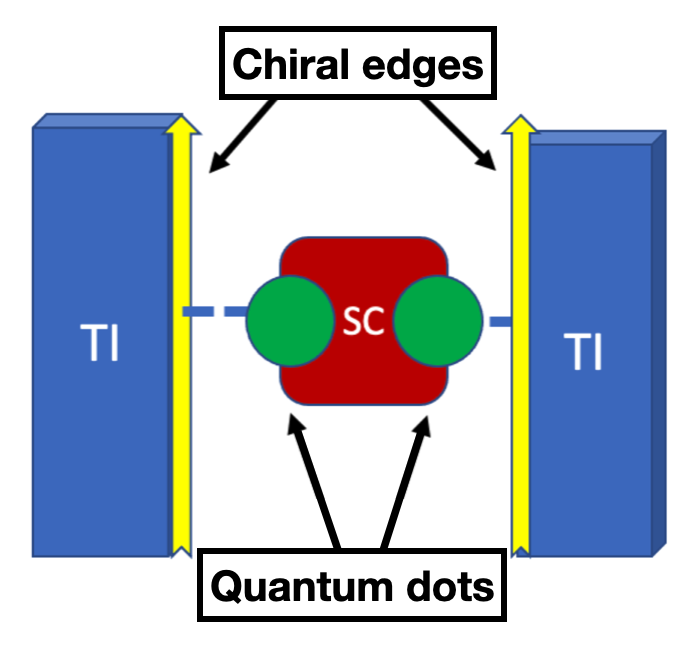}
    \caption{A sketch of the proposed Kondo device for $k=2$. The red rectangle is a superconducting island with charging energy $E_C$. The green dots are quantum dots with discrete single-particle spectrum of energy levels in proximity to the island. Electrons from the dots can tunnel to the chiral edges (in yellow) containing chiral electrons, and the chiral edges are supported by the appropriate topological insulators (in blue).}
    \label{fig:island}
\end{figure}

We repeat some details described in our previous publications \cite{PhysRevB.107.L201401,KoenigTsvelik2023} for the symplectic Kondo effect. The symplectic impurity spin is implemented by means of a superconducting island coupled to $k$ quantum dots:
\begin{equation}
\begin{split}
    \mathcal{H}_{\mathrm{dot}}&=E_C\left(2 N_C+n_d-N_g\right)^2\\
    &-\frac{\Delta}{2} \sum_{i=1}^k\sum_{\sigma\sigma'}\left[e^{-\ri \phi} d_{i \sigma}^{\dagger}\left[\sigma_y\right]_{\sigma \sigma^{\prime}} d_{i \sigma^{\prime}}^{\dagger}+ \text{H.c.}\right],
\end{split}
\end{equation}
where $n_d=\sum_{i=1}^k\sum_{\sigma} d_{i\sigma}^{\dagger} d_{i \sigma}$ is the total charge in the resonant zero energy states provided by the quantum dots and $N_C=-\ri\partial_{\phi}$ counts the number of Cooper pairs on the island. $E_C$ is the charging energy of the island, $N_g$ is the gate charge for the island, and $\Delta$ is the pairing amplitude. The labels $i=1,\ldots,k$ refer to the resonant zero energy states coupled to the superconducting island. The configuration of such a system with $k=2$ is shown in Fig. \ref{fig:island}.
It is physically realizable when Cooper pairs are converted into two electrons only locally at the $k$ resonant levels, which is satisfied when the distance between the quantum dots exceeds the superconducting coherence length. We will focus on the particle-hole invariant point $N_g=1$, where the Coulomb interaction on the quantum dot is irrelevant, and the impurity ground state manifold is spanned by generators of the $\mathfrak{sp}(2k)$ algebra. In the limit $0<E_C-\Delta\ll E_C,\Delta$, the ground state of the quantum dot is in the odd parity sector, $2k$-fold degenerate, and labelled by the channel and the spin indices of the singly occupied level $\ket{i,\sigma}$. For $t\ll E_C-\Delta$, the coupling to the leads
\begin{equation}
\label{tunn}
    \mathcal{H}_t=\sum_{i=1}^k\sum_{\sigma} \left[ t_i c^{\dagger}_{i\sigma}(x=0) d_{i\sigma}+\text {H.c.} \right] 
\end{equation}
is taken into account perturbatively by means of two virtual processes corresponding to transitions into one of the two lowest excited states with one additional and one missing charge. The addition of these two virtual processes leads to the following superexchange interaction:
\begin{equation}
\begin{split}
    \mathcal{H}_{\text{super}} &= \frac{2 |a\rangle \langle b|(t_{a'} c^\dagger_{a'}) (c_{b'}t_{b'}) [\delta_{ab'} \delta_{a' b} - (\sigma_y)_{ab}(\sigma_y)_{a'b'}]}{E_C-\Delta} \\ 
    & =\frac{2}{E_C-\Delta}\Big[t_a(c^{\dagger}_aT^A_{ab}c_b)t_b\Big]S^A, 
\end{split}
\end{equation}
where $a=(i,\sigma)$ is a multi-index including channel and spin indices.  $c_a (x)$ are fermionic annihilation operators, $T^A$ are generators of the Lie algebra $\mathfrak{g}=\mathfrak{sp}(2k)$ in the fundamental representation, and the impurity spin $S^A$ also transforms in the fundamental representation of $\mathfrak{g}=\mathfrak{sp}(2k)$. The resulting anisotropic superexchange interaction $g_{AB}J^A(0)S^B$ with $J^A(x) = \sum_{a,b} c_a^\dagger(x) T^A_{ab} c_{ b}(x)$ can be reduced to the diagonal form by a unitary transformation, and we obtain the symplectic Kondo model:
\begin{equation}
\label{eq:Kondo}
    \mathcal H = \ri v_F\int \rd x \sum_a c_a^\dagger\partial_x c_a + \sum_{A \in \mathfrak g} g_A J^A (0) S^A, 
\end{equation}
The terms of higher order in $t^2$ may contain powers of $S^A$, but they are irrelevant since they also contain higher powers of the fermionic operators. The exact solution exists for $t_i =t$, where $g_A = g =4t^2/(E_C-\Delta)$ and the model defined by (\ref{eq:Kondo}) possesses exact $Sp(2k)$ symmetry.

The fact that the current $J^A$ belongs to the $Sp_1(2k)$ theory can be checked by writing down the bulk Hamiltonian in terms of $4k$ species of Majorana fermions. The latter model is equivalent to the level-1 chiral $O(4k)$ Wess-Zumino-Novikov-Witten (WZNW) model. Then we use the conformal embedding \cite{conformalembedding,Kimura2021}:
\begin{equation}
    O_1(4k) = Sp_k(2) \oplus Sp_1(2k),
\end{equation}
to represent the Hamiltonian of free fermions as the sum of two WZNW Hamiltonians. It follows that Eq. (\ref{eq:Kondo}) can be written as a sum of two commuting parts:
\begin{equation}
\begin{split}
    \mathcal H &=\Big\{\mathcal{H}_{\text{WZNW}}[Sp_k(2)]\Big\} \\
    &+ \Big\{ \mathcal{H}_{\text{WZNW}}[Sp_1(2k)] + {\textstyle\sum}_{A\in\mathfrak{g}}g_AJ^A(0)S^A\Big\}.
\end{split}
\end{equation}
The central charge of the $Sp_1(2k)$ WZNW model is determined as
\begin{equation}
\label{eq:central}
c = \frac{k(2k+1)}{k+2}.
\end{equation}
We can check this by looking at a few simple examples: for $k=1$,  $Sp(2) = SU(2)$ and the central charge is 1, so we have $Sp_1(2) = SU_1(2)$; for $k=2$, $Sp(4) = O(5)$ and the central charge is 5/2 (five Majorana fermions), so we have $Sp_1(4) = O_1(5)$.

As we have mentioned above, the Kondo model defined in Eq. (\ref{eq:Kondo}) can be exactly solved only for $g_A =g$ where there is perfect symplectic symmetry \cite{KoenigTsvelik2023}, and then its low energy properties can be derived accordingly. We will review these results in Sec. \ref{sec:integrability}, and here we want to mention that the most important observation will be that the model possesses non-Abelian anyons in its ground state, which can be utilized to realize topological quantum computations.

In real situations there are always perturbations. Consider, for example, perturbations  which break the degeneracy of the energy levels provided by the quantum dots. The most likely perturbation  comes from the impact of a local chemical potential:
\begin{equation}
\label{pertD}
    \delta \mathcal{H} =\mu_i\sum_{\sigma}d^+_{i\sigma}d_{i\sigma},
\end{equation}
where $i=1,\ldots,k$ label the quantum dots. This perturbation, however, has zero overlap with the generators of the $\mathfrak{sp}(2k)$ algebra, since the latter satisfy
\begin{equation}
\label{eq:sp} 
 (S^A)^T = - \sigma^y S^A\sigma^y,
\end{equation}
where the Pauli matrix $\sigma^y$ acts on the spin indices, while the former does not satisfy Eq. (\ref{eq:sp}). Therefore, the perturbation in Eq. (\ref{pertD}) does not couple to the impurity spins directly, hence cannot contribute to relevant perturbations in the leading order in $\mu_i$, as was already discussed in Ref. \onlinecite{KoenigTsvelik2023}. Consequently, the presence of the symplectic symmetry will greatly weaken the influence of the level disorder, the latter being the bane of all anyon scenarios \bibnotemark[sela].

\bibnotetext[sela]{Through private communication with E. Sela we learned about preliminary numerical renormalization group data for the chiral Kondo chain model with $Sp(4)$ symmetry by M. Lotem, E. Sela \textit{et al.}, which suggests that both the level disorder in Eq. (\ref{pertD}) and the Kondo anisotropy in Eq. (\ref{eq:deltaga}) are relevant perturbations. This does not fit our conclusions supported by analytic calculations given in Appendix \ref{app:EKpoint}, where we have found only a single relevant direction.}

We may also imagine that in real situations the tunneling in Eq. (\ref{tunn}) will  be edge dependent. This will give rise to anisotropy of the Kondo couplings $g \rightarrow g_A$. Such anisotropy can also be generated  by higher order terms in the Shrieffer-Wolff transformation. Since the impurity is still described by the $\mathfrak{sp}(2k)$ generators, they can either modify couplings $g_A$ or generate irrelevant interaction terms with higher scaling dimensions. It is well known, however, that renormalization group flows, especially for for models with level $k=1$ Kac-Moody currents tend to converge, restoring the symmetry at low energies \cite{BalFisher,Symmetry}. In integrable models such as the anisotropic Kondo ones this effect can be traced beyond the perturbation theory. We have studied the case of strong anisotropy in the $k=2$ model following the approach pioneered by Emery and Kivelson \cite{EK1992}. Contrary to the common wisdom discussed above, we have found a single direction along which the model scales away from the symplectic quantum critical point and arrives at a new critical point corresponding to a Fermi liquid. More details are given at the end of Sec. \ref{sec:lowenergy} and in Appendix \ref{app:EKpoint}
\bibnotemark[sela].

\section{The Chiral Kondo Chain Model and its Integrability}
\label{sec:integrability}

\begin{figure}[htp!]
    \centering
    \includegraphics[width=\linewidth]{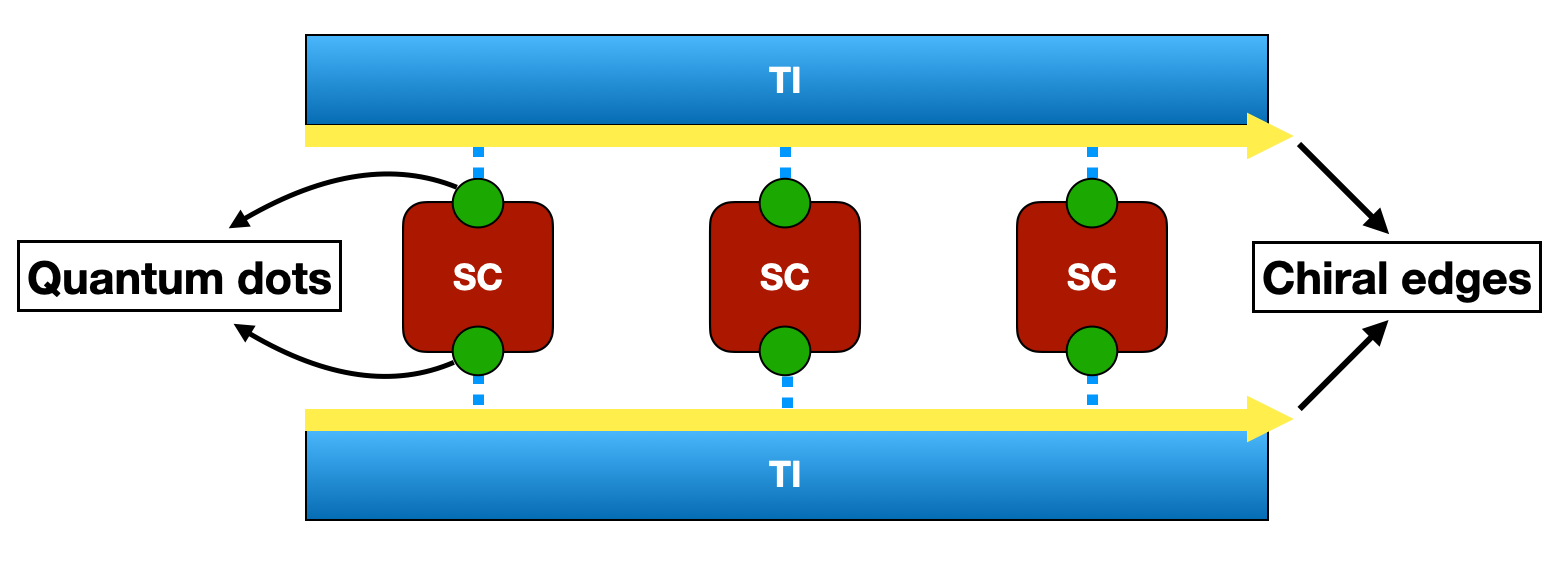}
    \caption{A sketch of the proposed chiral Kondo chain for $k=2$. The red rectangles are superconducting islands. The green dots are quantum dots. Electrons from the dots can tunnel to the chiral edges (in yellow) containing chiral electrons, which are supported by the topological insulators (in blue).}
    \label{fig:chainmodel}
\end{figure}

At the center of our project is the chiral Kondo chain model based on the symplectic Kondo model of Eq. (\ref{eq:Kondo}):
\begin{equation}
\label{eq:1Dmodel}
\begin{split}
    \mathcal H =& \ri v_F\int \rd x \sum_\alpha c_\alpha^\dagger\partial_x c_\alpha +  \sum_{j =1}^{N_s}g_j\sum_{A \in \mathfrak g} J^A (x_j) S^A(x_j), 
\end{split}
\end{equation}
where $\mathfrak g = \mathfrak{sp}(2k)$ and the single impurity spin is generalized to multiple impurity spins. The configuration of such a chain system with $k=2$ is shown in Fig. \ref{fig:chainmodel}. For simplicity, we will henceforth set the Fermi velocity $v_F = 1$. This type of models can be directly realized in topological electronic systems with gapped bulk where the boundary electronic modes are chiral, as discussed in Sec. \ref{sec:model}. In this section, we show that the chiral Kondo chain model defined in Eq. (\ref{eq:1Dmodel}) is integrable for an arbitrary number of impurity spins, which provides us a handle to analyze its low energy properties.

\begin{figure}[htp!]
    \centering
    \includegraphics[width=1\linewidth]{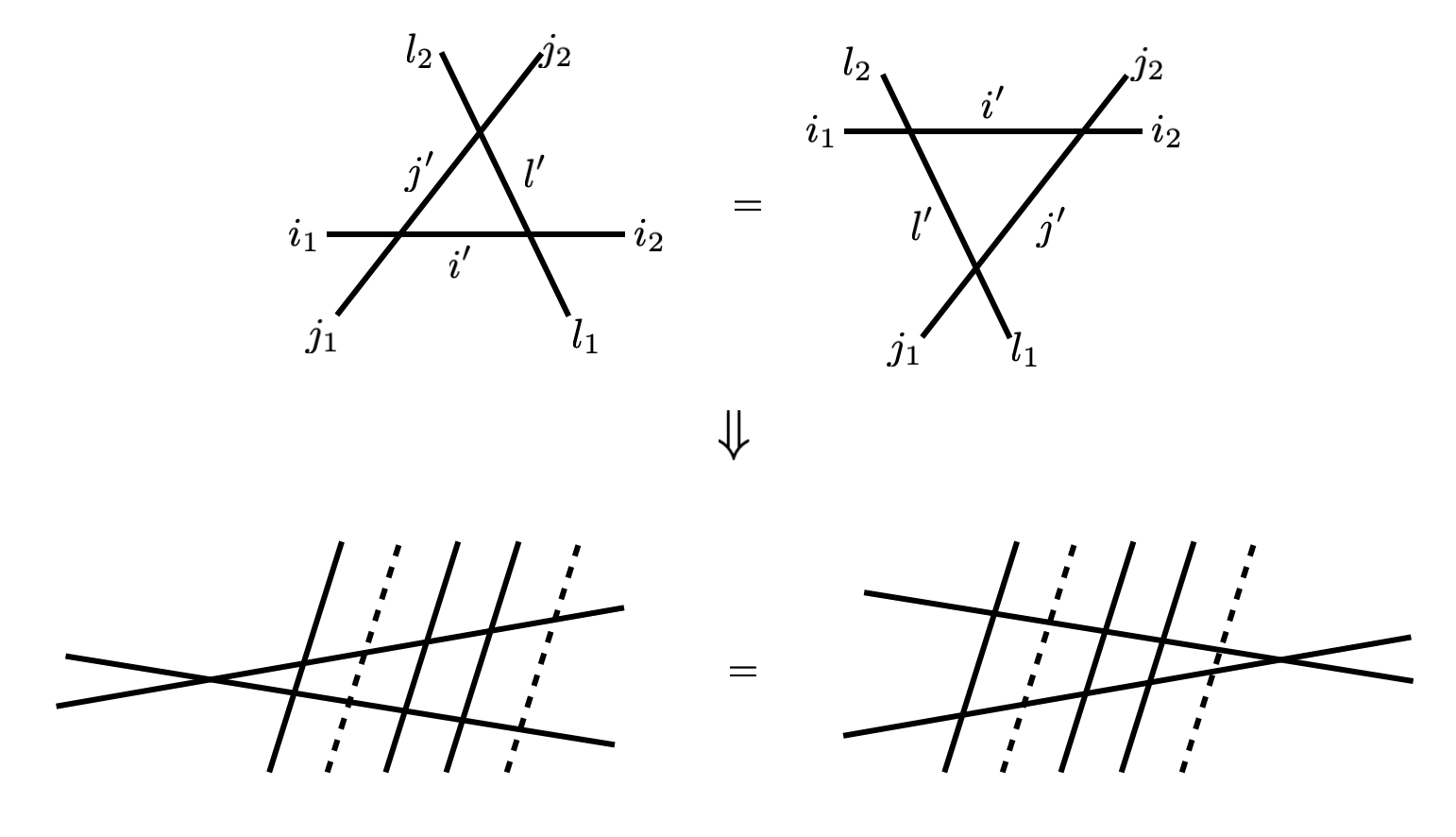}
    \caption{The Yang-Baxter equations for two-body scatterings, and the integrability for the chiral Kondo chain model. The line intersections stand for the scattering matrices. The lower panel illustrates the relation between the products of the monodromy matrices of the multi-impurity problem. In this panel, we differentiate the electrons and impurity spins by solid and dashed lines. The solid lines carry zero rapidity and the dashed ones carry rapidity $1/g_j$, where $g_j$ is the coupling for $j$-th impurity.}
    \label{fig:YB}
\end{figure}

The case of a single impurity spin $(N_s =1)$ and $k = 1$ was solved previously \cite{Andrei1980,Vigman1980,TsvelickWiegmann1983Review,AndreiLowenstein1983}. 
For chiral models it is straightforward to generalize the solution to the case with an arbitrary number of impurity spins and channels due to the link between the Kondo and chiral Gross-Neveau (GN) models, established in the pioneering paper by \citet{Andrei1980}. In this paper, the model defined by Eq. (\ref{eq:1Dmodel}) is treated as a limit of the chiral GN model, whose solution for an arbitrary number of right- and left-moving fermions $N_R,N_L$ was obtained in Ref. \onlinecite{AndreiLowenstein1979}. The difference between the Kondo chain model and the Lorentz invariant GN model is that in the former case the left-moving particles have zero velocity. Although in Ref. \onlinecite{Andrei1980} this method was used for the Kondo problem with $SU(2)$ symmetry, it is completely general and  holds for other Lie group symmetries. In particular, the solution for the symplectic symmetry can be extracted from  the results obtained for the chiral $Sp(2k)$ GN model \cite{PCF}.

Basically, the integrability of the model in Eq. (\ref{eq:1Dmodel}) lies in the fact that the $N$-body scattering problem for an arbitrary number of particles $N$ can be represented as a succession of two-body scattering problems. This is possible provided that eigenvalues are invariant under permutations of the momenta of the particles, which is fulfilled when the two-body $S$-matrices satisfy the Yang-Baxter equations shown in Fig. \ref{fig:YB}. For models with linear dispersions, the particle velocity acquires discrete set of values (in the given case it will be $v=1$ for the electrons and $v=0$ for the impurity). However, the integrability requires that the two-particle scattering matrices (known as $S$-matrices) $S(v,v)$ and $S(v,0)$ as functions of particle rapidities must lie on a continuous curve $S(\lambda)$ where the latter function satisfies the Yang-Baxter equations.

\begin{figure}[htp!]
	\centering
	\includegraphics[width=0.7\linewidth]{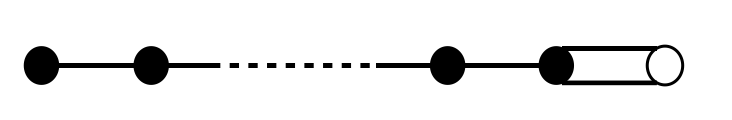}
	\caption{The Dynkin diagram for $\mathfrak{sp}(2k)$. The filled dots represent the long roots, the open dot represents the short root. For $k=2$, we only have one filled dot and one open dot connected by a double line.}
	\label{fig:Dynkin}
\end{figure}

For particles in the fundamental representation of the $Sp(2k)$ symmetry, the solution is \cite{Berg1978,Reshetikhin1985,Jimbo1986}:
\begin{widetext}
\begin{equation}
    \left[S_{\alpha\beta}^{\bar \alpha\bar \beta}\right](\lambda) = \frac{\left[\ri\lambda(\ri\lambda +k+1)\delta_{\alpha\beta}\delta_{\bar \alpha\bar \beta} +(\ri\lambda +k+1)\delta_{\alpha\bar \beta}\delta_{\beta\bar \alpha}- \ri\lambda \epsilon_\alpha\epsilon_\beta\delta_{\alpha,2k+1-\beta}\delta_{\bar \alpha,2k+1-\bar \beta}\right]}{(\ri\lambda +1)(\ri\lambda + k+1)},
\end{equation}
\end{widetext}
where $\epsilon_\alpha=1$ for $\alpha=1,\ldots,k$, and $\epsilon_\alpha=-1$ for $\alpha=k+1,\ldots,2k$, and we have adopted the notation that $\alpha = 1,\ldots,k$ correspond to multi-index $(i,-)$, and $\alpha = k+1, \ldots, 2k$ correspond to multi-index $(i,+)$ to make contact with the literature on integrable models. It indeed contains the solution of the Schr\"odinger equation for the model defined in Eq. (\ref{eq:1Dmodel}) as a particular case. Identifying $S(v,v') = S(\lambda = (v - v')/g_j)$  we obtain  
\begin{equation}
\begin{split}
    &\left[S_{\alpha\beta}^{\bar \alpha\bar \beta}\right](1,1)=\left[S_{\alpha\beta}^{\bar \alpha\bar \beta}\right](\lambda =0) = \delta_{\alpha\bar \beta}\delta_{\beta\bar \alpha}, \\
    &\left[S_{\alpha\beta}^{\bar \alpha\bar \beta}\right](1,0)=\left[S_{\alpha\beta}^{\bar \alpha\bar \beta}\right](\lambda =1/g_j)= \delta_{\alpha\beta}\delta_{\bar \alpha\bar \beta} \\
    & -\ri g_j\Big[\delta_{\alpha\bar \beta}\delta_{\beta\bar \alpha} - \epsilon_\alpha\epsilon_\beta\delta_{\alpha,2k+1 -\beta}\delta_{\bar \alpha,2k+1-\bar \beta}\Big] +O(g_j^2). 
\end{split}
\end{equation}
Equipped with the solution for the two-body scattering matrix, we can solve the many-body Schr\"odinger equation for particles on a ring of length $L =Na_0$ with periodic boundary conditions:
\begin{equation}
\label{eq:eigen}
    e^{\ri k_j L}\Psi = S(v_j, v_1)S(v_j,v_2)\cdots S(v_j,v_{N_R +N_L})\Psi,
\end{equation}
where $j=1,\ldots,N_R$, $\Psi$ is a suitably chosen component of the electron wave function, $N_R=N_e$ is the number of electrons, $N_L=N_s$ is the number of impurity spins, and $v=1,0$ for electrons and impurity spins respectively. The corresponding energy is $E =\sum_{j=1}^{N_R} k_j$. Using the general classification of  Eq. (\ref{eq:eigen}) for all simple Lie algebras given by Reshetikhin and Wiegmann \cite{PCF,RESHETIKHIN1987125,Reshetikhin1985}, we arrive at the following Bethe ansatz equations for $\mathfrak{g}=\mathfrak{sp}(2k)$ with $k=2$:
\begin{equation}
    \label{eq:driving1}
    \begin{split}
        & \left[e_2\left(\lambda_a^{(1)}\right)\right]^{N_e} \prod_{j=1}^{N_s}e_q\left(\lambda_a^{(1)}+1 / g_j\right)\\
        =& \prod_{\substack{b=1, b \neq a}}^{M_1} e_2\left(\lambda_a^{(1)}-\lambda_b^{(1)}\right) \prod_{c=1}^{M_2} e_{-2}\left(\lambda_a^{(1)}-\lambda_c^{(2)}\right) \\
        1=&\prod_{\substack{b=1, b \neq a}}^{M_2} e_4\left(\lambda_a^{(2)}-\lambda_b^{(2)}\right) \prod_{c=1}^{M_1} e_{-2}\left(\lambda_a^{(2)}-\lambda_c^{(1)}\right)
    \end{split}
\end{equation}
and with $k\geqslant 3$:
\begin{widetext}
\begin{equation}
\label{eq:driving2}
\begin{split}
    \left[e_2\left(\lambda_a^{(1)}\right)\right]^{N_e} \prod_{j=1}^{N_s} e_q\left(\lambda_a^{(1)}+1 /g_j\right) & = \prod_{\substack{b=1, b \neq a}}^{M_1} e_2\left(\lambda_a^{(1)}-\lambda_b^{(1)}\right) \prod_{c=1}^{M_2} e_{-1}\left(\lambda_a^{(1)}-\lambda_c^{(2)}\right), \\
    1& = \prod_{\substack{b=1, b \neq a}}^{M_i} e_2\left(\lambda_a^{(i)}-\lambda_b^{(i)}\right) \prod_{c=1}^{M_{i-1}} e_{-1}\left(\lambda_a^{(i)}-\lambda_c^{(i-1)}\right) \prod_{c=1}^{M_{i+1}} e_{-1}\left(\lambda_a^{(i)}-\lambda_c^{(i+1)}\right), \\
    1& = \prod_{\substack{b=1, b \neq a}}^{M_{k-1}} e_2\left(\lambda_a^{(k-1)}-\lambda_b^{(k-1)}\right) \prod_{c=1}^{M_{k-2}} e_{-1}\left(\lambda_a^{(k-1)}-\lambda_c^{(k-2)}\right) \prod_{c=1}^{M_k} e_{-2}\left(\lambda_a^{(k-1)}-\lambda_c^{(k)}\right), \\
    1& = \prod_{\substack{b=1, b \neq a}}^{M_k} e_4\left(\lambda_a^{(k)}-\lambda_b^{(k)}\right) \prod_{c=1}^{M_{k-1}} e_{-2}\left(\lambda_a^{(k)}-\lambda_c^{(k-1)}\right),
\end{split}
\end{equation}
\end{widetext}
where the involved notations are explained as follows. The index $i=1,\ldots,k$ enumerates the simple roots of the Lie algebra $\mathfrak{g}=\mathfrak{sp}(2k)$, including the long roots $i=1,\ldots,k-1$ and the short root $i=k$, as shown on the Dynkin diagram in Fig. \ref{fig:Dynkin}. 
The parameters $\lambda^{(i)}_a$ are called rapidities and $M_i$ is the number of rapidities corresponding to the $i$-th root. Each $\lambda^{(i)}$ interacts with the set of rapidities $\lambda^{(i')}$ of the same root $i'=i$ and of adjacent roots $i'=i\pm 1$, as shown on the Dynkin diagram in Fig. \ref{fig:Dynkin}. The function $e_n(x)$ is defined as
\begin{equation}
    e_n(x)=\frac{x+\ri n/2}{x-\ri n/2}
\end{equation}
and encodes the information of the scattering matrix. The subscript $n$ for adjacent roots is determined according to the lines of the Dynkin diagram in Fig. \ref{fig:Dynkin}: $n=-1$ if the roots are connected by a single line and $n=-2$ if they are connected by a double line. The subscript for the same root is positive and determined by its length: $n=2$ for the long roots and $n=4$ for the short root. The driving term on the left hand side of the first equation in (\ref{eq:driving1}) and (\ref{eq:driving2}) includes two factors, where the first factor has subscript $2$. If the subscript of the second factor $q=2$, we have the usual Fermi liquid ground state. Here we have $q=1$ instead, which corresponds to the quantum critical ground state. The position of the driving term among the Bethe ansatz equations (labelled by $m$) is determined by the condition $l=2(m-1)+q$, where $l$ is the occupation number of the orbital degree of freedom. Since we have a single orbital for each species of fermions, we set $l=1$, leading to $m=1$, thus the driving term is placed on the left hand side of the first Bethe ansatz equation.

Notably, the above formulae are written for the general case where impurities have different isotropic couplings. The exact solution tolerates this disparity provided the symplectic symmetry is not violated. The net result of this is that instead of a single Kondo temperature we will have a certain distribution which will not affect the low energy behavior \cite{PhysRevB.107.155417}.

\section{Low Energy Properties of the Chiral Kondo Chain Model}
\label{sec:lowenergy}
  
From the viewpoint of thermodynamics, the difference between the case with a single impurity and the case with a multi-impurity chain is minimal: the effect of impurities is additive, and the corresponding part of the free energy for the chain is just multiplied by the factor $N_L=N_s$. The exact solution  provides  us with an additional handle to manipulate the system: we do not need to restrict ourselves to the quantum critical point, as was done in literature \cite{Sela1,Sela2}. However, the main result is in agreement with these papers. Specifically, the ground state entropy is $S(T=0) = N_s \ln d_k$, where  
\begin{equation}\label{eq:DegeneracySummarySec}
    d_k = 2\cos\left(\frac{\pi}{k+2}\right),
\end{equation}
is the anyonic quantum dimension \cite{PhysRevLett.67.161}. This indicates that in the low energy limit the symplectic chiral Kondo chain can be described as a set of anyons. These anyons are localized on the impurity sites in the sense that various local operators related to impurities such as impurity spins (see, for instance, Eq. (\ref{eq:mutation}) below) are expressed in terms of these anyons. Below we will first summarize the main results for a single impurity obtained from the exact solution. Then we will discuss the connection to the conformal field theory (CFT) approach and identify the operator content at the QCP for further developments.

\subsection{Thermodynamics from Exact Solution}

We summarize the main results for a single impurity obtained from the exact solution \cite{KoenigTsvelik2023}. The generalization to the chiral Kondo chain is simply an additive effect. In the absence of external magnetic fields, the free energy is
\begin{equation}
\label{eq:freeenergy}
\begin{split}
    F \stackrel{T \ll T_K}{\simeq}& -\frac{\pi c}{12} \frac{T^2}{E_F}+\frac{1}{N_e}\left[-T \ln \left(d_k\right)-\frac{\pi c_k}{12} \frac{T^2}{T_K}-\cdots\right.\\
    &\left.-\text {const.} \times T\left(\frac{T}{T_K}\right)^{k+1} \ln \left(T_K / T\right)\right]
\end{split}
\end{equation}
where $T_K \sim E_F \exp\left(-\frac{\pi v_F}{(k+1)g}\right)$ is the Kondo temperature, $c_k=c / \left[2 \cos \left(\frac{\pi }{2 k+2}\right)\right]$ with $c$ the central charge for $Sp_1(2k)$ theory in Eq. (\ref{eq:central}). The impurity free energy in the square bracket contains a zero temperature entropy consistent with the anyonic quantum dimension in Eq. (\ref{eq:DegeneracySummarySec}). This ground state degeneracy is the same as that for the $k$-channel $SU(2)$ Kondo effect \cite{Sela1}. However, different from the $k$-channel $SU(2)$ Kondo effect, the leading finite temperature corrections are Fermi liquid like. A similar behavior is also true for the zero temperature impurity magnetization in the presence of a small local Zeeman field $H_i\ll T_K$, which takes the form
\begin{equation}
\label{eq:Hfield}
\begin{split}
    M_i=&\alpha_1 H_i / T_K+\alpha_3\left(H_i / T_K\right)^3+\cdots\\
    &+\beta\left(H_i / T_K\right)^{k+1} \times \begin{cases}1, & k \text { odd } \\ \ln \left(T_K / H_i\right), & k \text { even }\end{cases}
\end{split}
\end{equation}
where $\alpha_{1,3,\ldots}$ and $\beta$ are dimensionless constants, and $i=1,\ldots,k$ label the quantum dots. The fact that the leading order corrections for low energy thermodynamics are Fermi liquid like has important implications on the operator content at the QCP, which we will discuss below.

\subsection{The CFT Analysis}
\label{sec:cft}

The CFT analysis of the $Sp_1(2k)$ theory for a general $k$ is contained in Ref. \onlinecite{Kimura2021}, and the special case of $k=2$ ($Sp(4)\simeq SO(5)$) is discussed in Ref. \onlinecite{PhysRevB.103.195131,PhysRevLett.126.147702} via the Abelian bosonization technique. Here we will summarize the main points and make connection to the results obtained from exact solutions.

The $Sp_1(2k)$ WZNW theory has $k$ primary fields $\Phi_a$ with scaling dimensions \cite{BOUWKNEGT1999501} 
\begin{equation}
h_a = \frac{a(2k+2-a)}{4(k+2)}, \quad a=1,\ldots, k,
\end{equation}
transforming according to the  representations $\Lambda_a$ of the Lie algebra $\mathfrak{sp}(2k)$ with dimensionalities
\begin{equation}
\label{eq:dim}
    \operatorname{dim}\Lambda_1=2k, \quad \operatorname{dim}\Lambda_a=C_{2k}^a - C_{2k}^{a-2}, \quad a\geqslant 2,
\end{equation}
where $C^a_{2k}=(2k)!/[a!(2k-a)!]$ is the binomial coefficient. The fusion rules for $Sp_1(2k)$ primaries can be expressed as \cite{WALTON1990777,Cummins_1991}
\begin{equation}
\label{eq:fusionrule}
\begin{split}
    a \otimes a' = & |a-a'|, ~|a-a'|+2, ~\ldots, \\
    & \operatorname{min}\{a+a', 2k-a-a'\},
\end{split}
\end{equation}
where $a=0,1,\ldots,k$, including the vacuum. For $k=1$, there is only one primary with $h=1/4$, which is obviously the $SU(2)$ matrix field with entries $\exp[\ri\sqrt{2\pi}\phi]$, $\exp[\ri\sqrt{2\pi}\theta]$, where $(\phi,\theta)$ is a set of conjugate variables. For $k=2$, there are two primaries, one with $h_1 = 5/16$ being a product of five Ising model conformal blocks, and the other with $h_2 = 1/2$ being a Majorana fermion.

To figure out the properties of the QCP, we follow the approach of Ref. \onlinecite{AFFLECK1990517,AFFLECK1991849}, where the vacuum is redefined by absorbing the impurity such that the description of the system is simplified. The WZNW Hamiltonian of the $Sp_1(2k)$ theory can be expressed in the Sugawara form \cite{review1,review2}:
\begin{equation}
\begin{split}
    & \mathcal{H}_{\text{bulk}} = \frac{2\pi}{k+2} \sum_{n \geq 0}\sum_{A\in\mathfrak{g}} J^A_{-n}J^A_n, \\
    & J_n^A = \frac{1}{2\pi}\int_{-L/2}^{L/2}\rd x ~J^A(x)e^{2\ri\pi xn/L},
\end{split}
\end{equation}
where the current operator satisfies the following Kac-Moody algebra:
\begin{equation}
    [J^A_n,J_m^B] = \ri f^{ABC}J^C_{n+m} + \frac{1}{2}n\delta_{AB}\delta_{n+m,0}. \label{eq:KacMoody}
\end{equation}
Then the Kondo term can be absorbed into the Hamiltonian through the redefinition of the currents \cite{AFFLECK1991849,AFFLECK1990517}:
\begin{equation}
{\cal J}_n^A = J^A_n + (k/2 +1)g S^A,
\end{equation}
such that we can ``complete the square" in the expression of the Hamiltonian \cite{AFFLECK1990517}. The QCP is obtained by requiring that the Kac-Moody algebra of Eq. (\ref{eq:KacMoody}) remains unchanged, and this is achieved at $g(k/2+1) =1$. This absorption leads to the redefinition of the vacuum through the fusion with the impurity spin, the spin being a $2k\times 2k$ matrix corresponding to the primary field $\Phi_1$. This is known as the ``fusion ansatz", which we briefly review in Appendix \ref{app:primarycorrelation}. As a result of such fusion the impurity spin operators experience ``transmutation" - at the critical point they are expressed in terms of the bulk fields located at the impurity sites. Our goal now is to establish this correspondence for the symplectic symmetry.

The distinct feature of the $Sp_1(2k)$ theory is the high scaling dimensions of the irrelevant operators perturbing the QCP \cite{PhysRevB.107.L201401}. This is consistent with the exact solution results in Eq. (\ref{eq:freeenergy}) and (\ref{eq:Hfield}), where the non-analytic terms appear in high orders of the expansion in $T/T_K$ and $H_i/T_K$. This is unlike the $SU(2)$ Kondo effect and, as we will show later, constitutes an advantage of the symplectic Kondo effect as a platform for quantum information applications. 

 To elaborate on this point, we consider the case $k=2$ where the CFT consists of five Majorana fermions ($Sp_1(4)$ = $O_1(5)$) and the two primaries are a product of five Ising model conformal blocks and a Majorana fermion. The ground state entropy suggests the existence of a Majorana zero mode. From five Majorana fermions $\chi_1,\ldots,\chi_5$ and one Majorana zero mode $\xi$, we can construct the perturbation~\cite{PhysRevB.103.195131}
\begin{equation}
\delta \mathcal H \sim \xi \chi_1(0)\cdots\chi_5(0),
\end{equation}
which generates the non-analytic correction to the specific heat in Eq. (\ref{eq:freeenergy}):
\begin{equation}
    \delta C_V \sim (T^3/T_K^2)\ln(T_K/T).
\end{equation}
From Eq. (\ref{eq:Hfield}) we can see that the magnetic field $H_i$ is coupled to the operator with scaling dimension $\Delta =3/2$ \cite{PhysRevB.103.195131,PhysRevB.107.L201401}, which generates the non-analytic correction to the magnetization 
\begin{equation}
\label{eq:k2}
\delta M_i \sim (H_i/T_K)^3\ln(T_K/H_i). 
\end{equation}
The correction in Eq. (\ref{eq:k2}) suggests that for $k=2$ the non-analytic contribution to the impurity spin comes from the operator containing three Majorana fermions, that is from a descendant of the primary field $\Phi_2$. We generalize this conjecture for general $k$ (cf.~\onlinecite{Sela1} for the case $SU(2)_k$) and suggest that the transmuted impurity spin is given by the sum of the $Sp_1(2k)$ current which generates the Fermi liquid like corrections and the descendant of the primary field $\Phi_2$ which generates the non-analytic corrections:
\begin{equation}
\label{eq:mutation}
    S_{\text{imp},j} = J(x_j) + \Gamma_j J_{-1}\Phi_2(x_j)+\cdots,
\end{equation}
where the dots stand for less relevant terms, and we introduce the impurity site label $j$ for the generalization to the multi-impurity case. The second term is a product of an operator $\Gamma$ acting on the anyonic Hilbert space and a descendant of the primary field $\Phi_2$ with scaling dimension $h_2$. The primary field $\Phi_2$ is chiral and has a noninteger scaling dimension. As a result, the pair correlation function of $J_{-1}\Phi_2$ at the same site behaves as $\tau^{-2-2h_2}$. However, the correlation function of the physical spin must be even in time, which fixes the time dependence of the correlation function of $\Gamma$:
\begin{equation}
\begin{split}
    & \lrangle{\Gamma_j(\tau)\Gamma_j(0)}\sim e^{\ri \pi h_2\operatorname{sgn}\tau}\\
    \Rightarrow~ &  \lrangle{S^A_{\text{imp},j}(\tau)S^B_{\text{imp},j}(0)} \sim \delta^{AB}\left( \frac{1}{|\tau|^2}+\frac{\Xi}{|\tau|^{2+2h_2}}\right),
\end{split}
\end{equation}
where $\Xi$ is a constant coefficient. On the other hand, the anyonic property of the operator $\Gamma$ is determined by its fusion rules, and the latter ones point to its  correspondence to the primary field $\Phi_1$ in the fusion ansatz. To get an insight we consider two relevant cases $k=2$ and $k=3$, using the $Sp_1(2k)$ fusion rules in Eq. (\ref{eq:fusionrule}). For $k=2$, we have $\Phi_a$ with $a=0,1,2$ and the fusion rules for them are
\begin{equation}
    0\times a=a, ~ 1\times 1=0+2, ~ 1\times 2=1, ~ 2\times 2=0.
\end{equation}
Therefore we have the following identification with the system of Ising anyon \cite{NayakDasSarma}:
\begin{equation}
\label{eq:k2Ising}
    0\to \mathbf{1}, \quad 1\to \sigma, \quad 2\to \psi,
\end{equation}
and $\Gamma$ plays the role of $\sigma$, the Ising anyon. This is also in agreement with the anyonic quantum dimension $d_2=\sqrt{2}$ from the ground state entropy. For $k=3$, we have $\Phi_a$ with $a=0,1,2,3$ and the fusion rules for them are
\begin{equation}
    0\times a = a, ~ 3\times a=(3-a), ~ 1\times 1=2\times 2=0+2, ~ 1\times 2=1+3.
\end{equation}
Here the set of rules $3\times a=(3-a)$ plays the special role of the automorphism of the fusion algebra, which maps $3$ to $0$ and $1$ to $2$, and we have the following identification with the system of Fibonacci anyon \cite{10.1143/PTPS.176.384,PRXQuantum.2.010334}:
\begin{equation}
\label{eq:k3Fibonacci}
    0,3\to \bm{1}, \quad 1,2\to \tau.
\end{equation}
Consequently, $\Gamma$ plays the role of $\tau$, the Fibonacci anyon \bibnote{We note that $\Phi_1$ and $\Phi_2$ are primary fields, while $\Gamma$ is the Fibonacci anyon. They share the same fusion rules but they are not identified with each other.}, whose properties are briefly reviewed in Appendix \ref{app:Fibonacci}. Again, this is in agreement with the anyonic quantum dimension $d_3=(1+\sqrt{5})/2$ from the ground state entropy. 

Equation (\ref{eq:mutation}) helps to shed light on the problem of stability of the QCP with respect to the anisotropy of the Kondo exchange interaction. Indeed, if we add a perturbation 
\begin{equation}
\label{eq:deltaga}
\delta \mathcal{H} = \sum_j \delta g_{AB} \cdot J^A(x_j)S^B(x_j),    
\end{equation}
then, after ``completing the square'' and replacing $S^A(x_j)$ by Eq. (\ref{eq:mutation}) at criticality, we find that the leading term of Eq. (\ref{eq:deltaga}) is irrelevant for the single impurity and marginal for the chain. Hence the coupling to the anyons does not emerge in the leading order in $\delta g$. In fact, as it follows from the analysis of the case $k=2$ presented in Appendix \ref{app:EKpoint}, for a certain type (not all types) of Kondo exchange anisotropy, the effect emerges in the second order, making the corresponding perturbation relevant \bibnotemark[sela].

\section{Extraction of Information from the Correlation Functions}
\label{sec:correlation}

The anyons residing at the impurity sites, which are supported by the chiral Kondo chain model with symplectic symmetry, can be utilized for measurement-only quantum computations \cite{BondersonNayak2008}. The essential ingredients of the measurement-only quantum computation protocol are briefly reviewed in Appendix \ref{app:MBQC}, where we can see that the building blocks of the protocol are the projective measurements of the fusion channel of anyons. Such projective measurements can be carried out by measuring the correlation functions of the chiral Kondo chain model. The general theory for such correlation functions is described in Ref. \onlinecite{Sela2} and in this section we specialize it to the symplectic case that we have been considering.

Information of the total fusion channel of all anyons between two spatial points can be extracted from the correlation function of the $Sp_1(2k)$ primary fields, the latter can be obtained under the multi-fusion ansatz \cite{Sela2} and is briefly reviewed in Appendix \ref{app:primarycorrelation}. On the other hand, for the purpose of quantum computations, what is needed is the fusion channel of two arbitrary anyons, which can be extracted from the pair correlation function of the impurity spins. We have conjectured that the impurity spin immersed in the chiral fermions gets decorated by an operator $\Gamma_j$ acting on the anyonic Hilbert space, such that their correlation encodes the information of the fusion channel of the two anyons associated with the two impurity sites. In case of the $Sp_1(2k)$ theory, the conjecture takes the form in Eq. (\ref{eq:mutation}), so for $t_i>t_j$ we have
\begin{equation}
\label{eq:correlationS}
\begin{split}
    &\left\langle S^{A}_{\text{imp},i}(t_i) S_{\text{imp},j}^{B}(t_j)\right\rangle_{a_{ij}} \sim \\
    &\delta^{AB} \left\{ \frac{1}{(t_{ij} - x_{ij} +\ri 0)^2} + \frac{\mathcal{F}_k\left(a_{ij}\right)
    }{\left(t_{ij}-x_{ij} +\ri 0\right)^{2+ \frac{2k}{k+2}}}\right\},
\end{split}
\end{equation}
where $t_{ij}=t_i-t_j, x_{ij}=x_i-x_j$, $a_{ij}$ denotes the fusion channel of the two anyons, and $\mathcal{F}_k(a_{ij})$ encodes the dependence on $a_{ij}$. Here the subscript $i,j$ label the impurity sites. The task remains is to find the expression for $\mathcal{F}_k(a_{ij})$, or at least, to show the dependence of $\mathcal{F}_k(a_{ij})$ on $a_{ij}$. 

We here look at two relevant cases $k=2$ and $k=3$ for illustrations. For $k=2$, $\Phi_2$ has scaling dimension $1/2$ and corresponds to the Majorana fermion. Consequently, $\Gamma_i$ corresponds to the Majorana zero mode $\xi_i$, with the commutation relations of the Clifford algebra:
\begin{equation}
  \{\xi_i,\xi_j\} = \delta_{ij}, ~~ \xi^2_i = 1/2.  
\end{equation}
On the other hand, we have identified $\Gamma$ with the Ising anyon in Sec. \ref{sec:cft}, from the fusion rules of $\Phi_1$. The two conclusions are in indeed consistent, where the Majorana zero mode is identified with the Ising anyon, and the bi-anyonic fusion channels can be constructed as
\begin{equation}
    \begin{split}
        & f_{ij}=\frac{\xi_i+\ri \xi_j}{\sqrt 2}, ~ f^{\dagger}_{ij}=\frac{\xi_i-\ri \xi_j}{\sqrt 2} ~ \Rightarrow ~ \ri \xi_i\xi_j=f^{\dagger}_{ij}f_{ij}-1/2; \\
        & \bm{1}\to \ket{0}, \quad \psi \to f^{\dagger}_{ij}\ket{0}.
    \end{split}
\end{equation}

We can see that $2\ri\xi_i\xi_j=\mathcal{P}_{ij}$ is the parity operator, and the fusion channels $\ket{a_{ij}}=\ket{0}, \ket{a_{ij}}=f^{\dagger}_{ij}\ket{0}$ are its eigenstates with eigenvalues $-1$ and $+1$ respectively. Hence a measurement of the impurity spin correlation function of Eq. (\ref{eq:correlationS}) is indeed a projective measurement of the bi-anyonic fusion channel, with $\mathcal{F}_2(\bm{1})\sim \ri$ and $\mathcal{F}_2(\psi)\sim -\ri$. The essence is that $\mathcal{F}_2(\bm{1})$ and $\mathcal{F}_2(\psi)$ differ by a sign.

Alternatively, we infer the behaviors of $\mathcal{F}_k(a_{ij})$ in another way that can be generalized to $k>2$. Since $\Gamma$ share the same fusion rule with $\Phi_1$, it is natural to assume that their correlation functions share the same dependence on the fusion channel. The correlation functions of $\Phi_1$ with two impurity spins in between is obtained in Appendix \ref{app:primarycorrelation} as
\begin{equation}
\label{eq:phi1phi1}
    \lrangle{\Phi_1(z_i)\Phi_1(z_j)}\sim \frac{1}{(z_i-z_j)^{2h_1}}\frac{S^1_{c_{ij}}/S^1_0}{S^0_{c_{ij}}/S^0_0},
\end{equation}
where $z\equiv \tau-\ri x$ is the complex coordinate, and $c_{ij}$ takes the value 0 or 2, which comes from the fusion rule $1\times 1=0+2$. The modular $S$-matrix element $S^{a'}_a$ for the $Sp_1(2k)$ theory is given by
\begin{equation}
    S_a^{a^{\prime}}=\sqrt{\frac{2}{2+k}} \sin \left[\frac{\pi(a+1)\left(a^{\prime}+1\right)}{2+k}\right].
\end{equation}
Consequently, we can extract from Eq. (\ref{eq:phi1phi1}) the fusion channel-dependent part, and assign it to $\mathcal{F}_k(a_{ij})$:
\begin{equation}
\label{eq:Faa}
    \mathcal{F}_k(a_{ij}) = \frac{S^1_{c_{ij}}/S^1_0}{S^0_{c_{ij}}/S^0_0},
\end{equation}
where the $\mathcal{F}_k(a_{ij})$ is normalized such that $\mathcal{F}_k(\bm{1})= 1$ ($c_{ij}=0 \to \bm{1}$, the vacuum). For $k=2$, we have the fusion channels $c_{ij}=0 \to a_{ij}=\bm{1}$ and $c_{ij}=2 \to a_{ij}=\psi$, as can be seen from Eq. (\ref{eq:k2Ising}), then Eq. (\ref{eq:Faa}) gives us
\begin{equation}
    \mathcal{F}_2(\bm{1}) = 1, \quad \mathcal{F}_2(\psi) = -1.
\end{equation}
This is essentially the same as that obtained from the Majorana zero mode algebras, where $\mathcal{F}_2(\bm{1})$ and $\mathcal{F}_2(\psi)$ differ by a sign.

For $k=3$, $\Phi_2$ has scaling dimension $3/5$ and we do not know \textit{a priori} its zero mode $\Gamma$. However, we have identified $\Gamma$ with the Fibonacci anyon $\tau$ in Sec. \ref{sec:cft}, from the fusion rules of $\Phi_1$. We are not able to calculate the correlation function of the Fibonacci anyons in case of coincident positions, so we resort to Eq. (\ref{eq:Faa}). By identifying $c_{ij}=0 \to a_{ij}=\bm{1}$ and $c_{ij}=2 \to a_{ij}=\tau$ from Eq. (\ref{eq:k3Fibonacci}), we obtain
\begin{equation}
    \mathcal{F}_3(\bm{1})= 1, \quad \mathcal{F}_3(\tau) = \frac{1}{2}(-3+\sqrt{5}).
\end{equation}
Different from the case of $k=2$, now $\mathcal{F}_3(\bm{1})$ and $\mathcal{F}_{3}(\tau)$ differ by sign and modulus, which should be easier to detect experimentally.

\begin{figure}[htp!]
	\centering
	\includegraphics[width=0.85\linewidth]{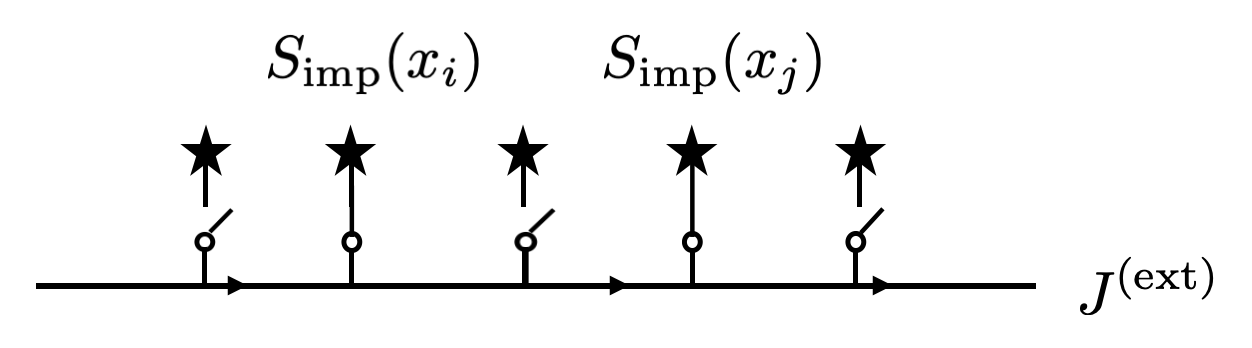}
	\caption{Measurement of the impurity spin (denoted as the star in the figure) correlation function through an extra single channel of chiral current with switches.}
 \label{fig:measure}
\end{figure}

A possible measurement of the correlation function of the impurity spins can be done by locally coupling the two impurity spins to an extra single channel of chiral current with switches, as shown in Fig. \ref{fig:measure}, and measuring the corresponding conductance at finite frequency.
Different from the channels responsible for creating the degenerate ground states, this extra single channel is controlled by the switches such that each time it couples only to those two impurity spins, whose correlation function is to be measured. Later in Sec. \ref{sec:error}, we will argue that this extra current does not effect the original anyonic structure of the system. 

To couple to the impurity spins, the extra current must 
belong to the $\mathfrak{sp}(2k)$ Kac-Moody algebra. This can be achieved by various means, for instance, one can use spin polarized current. The following discussion does not depend on the specific arrangement, we just assume that the extra current transforms as some generator of the $\mathfrak{sp}(2k)$ algebra and hence by symmetry can be coupled to at least one projection of the impurity spins, say $A$. Then the Hamiltonian at criticality for the whole system including the extra current is
\begin{equation}
\label{eq:local}
\begin{split}
    & \mathcal{H}_{\text{tot}} = \mathcal{H} +\pi \int \rd x ~J^{\text{(ext)}}(x)J^{\text{(ext)}}(x)+\lambda\delta \mathcal{H}_{ij}, \\
    & \delta \mathcal{H}_{ij} = J^{(\text{ext})}(x_i)\cdot S^A_{\text{imp},i}+J^{(\text{ext})}(x_j)\cdot S^A_{\text{imp},j},
\end{split}
\end{equation}
where $\mathcal{H}$ is the Hamiltonian for the original system, and $\lambda$ is the coupling constant. The corresponding current operator is determined by
\begin{equation}
\begin{split}
    & I^{(\text{ext})}=\frac{\rd N^{(\text{ext})}}{\rd t}=-\ri[N^{(\text{ext})},\mathcal{H}_{\text{tot}}], \\ 
    & N^{(\text{ext})}=\int^{x_0} \rd x~J^{\text{(ext)}}(x).
\end{split}
\end{equation}
Using the chiral anomaly
\begin{equation}
    [J^{\text{(ext)}}(x),J^{\text{(ext)}}(x')]=\frac{\ri}{2\pi}\delta'(x-x'),
\end{equation}
the current operator can be expressed as
\begin{equation}
\label{eq:lambdacurrent}
\begin{split}
    I^{(\text{ext})}(x_0)=J^{(\text{ext})}(x_0)+\frac{\lambda}{2\pi}&\left[\delta(x-x_i)S^A_{\text{imp},i}\right. \\
    & \left. + \delta(x-x_j)S^A_{\text{imp},j} \right].
\end{split}
\end{equation}
Then the conductance is related by the Kubo formula to the correlation function of this particular spin component placed at impurity site $i$ and $j$. The part of the finite frequency conductance that depends on the fusion channel of the anyons is
\begin{equation}
\label{eq:conductance}
    \delta G_{ij}(\omega)\propto \lambda^2 \omega^{\frac{2k}{2+k}} \mathcal{F}_k(a_{ij})\cos(\omega x_{ij}),
\end{equation}
where $x_{ij}=x_i-x_j$. Hence by measuring the conductance at finite frequency, one can effectively measure the bi-anyonic fusion channel. This idea lies at the foundation of our proposal for the topological quantum computation discussed in detail in Sec. \ref{sec:TQC}. Here we have only showed that $G_{ij}(\omega)$ will take different values for different fusion channels, without pinning down the exact values. This is enough from a pragmatic viewpoint - the association of experimentally measured $G_{ij}(\omega)$ with the specific fusion channel will be individually determined through the calibration process in actual implementations. Also, as will be discussed in Sec. \ref{sec:error}, the perturbation introduced by $\delta\mathcal{H}_{ij}$ in Eq. (\ref{eq:local}) is irrelevant, so the measurement, although being non-topological, does not affect the type of anyons supported by the chiral Kondo chain.

\section{Topological Quantum Computing Protocol}
\label{sec:TQC}

Currently, there are  four main different approaches to the quantum computing protocols: the quantum circuit model \cite{nielsen_chuang_2010}, the measurement-based quantum computation \cite{PhysRevLett.86.5188,PhysRevA.68.022312}, the adiabatic quantum computation \cite{farhi2000quantum,RevModPhys.80.1061}, and the topological quantum computation \cite{FaultTolerant,NayakDasSarma,PachosBook2012,10.21468/SciPostPhys.3.3.021}. Compared with the other approaches, the topological quantum computation has the advantage of building the fault-tolerance already on the level of hardware, utilizing the anyonic ground state manifold with topological degeneracy \cite{KITAEV20062}. Conventionally, the topological quantum computation is performed by adiabatic braiding of anyons, which is prone to uncontrollable error channels such as accidental braiding with unwanted anyons and thermal excitation of unaccounted anyons. However, these flaws can be  mitigated by combining the topological quantum computation with measurement-only quantum computation \cite{BondersonNayak2008}, where the adiabatic braiding of anyons is replaced by repeated projective measurements of the bi-anyonic fusion channels of anyons. Such a protocol is particularly suitable for the quantum computer based on the chiral Kondo chain model considered here, since the local operators are expressed in terms of anyons at the impurity sites.

In the previous sections, we have shown that the chiral Kondo chain model with $Sp(2k)$ symmetry possesses non-Abelian anyons in its ground state. For $k=2$, we have Ising anyons, and for $k=3$, we have Fibonacci anyons. In the following discussion we will focus on the case $k=3$ with Fibonacci anyons which support universal quantum computations \cite{NayakDasSarma}. As the first step we will propose the microarchitecture of the quantum computer based on the 3-channel chiral Kondo chain, and then consider the choice of materials for its realization. Finally we will discuss the sources and effects of errors on the proposed quantum computer.

\subsection{Microarchitecture of the Quantum Computer}

Let us summarize our results. The Fibonacci anyons created by coupling the superconducting islands to three (pseudo)spin degenerate chiral channels are localized on the superconducting islands. At criticality, these Fibonacci anyons are decoupled from the rest of the system. We suggest to use these immobile anyons for topological quantum computation through the measurement-only quantum computing protocol \cite{BondersonNayak2008}. The essential idea is that the physical transport of an anyon can be replaced by the quantum teleportation of the information encoded in the ground state, and such quantum teleportation can be realized by repeated projective measurements of the bi-anyonic fusion channel among  two given anyons and an extra auxiliary anyon. To promote the quantum teleportation between two anyons into the braiding of them, we require a further extra auxiliary anyon. The corresponding details are reviewed in Appendix \ref{app:MBQC}. Consequently, the realization of the braiding requires a total number of four anyons, two of which are auxiliary. If the measurements of the bi-anyonic fusion channel can be done nondemolitionally, as we argue below, the equivalent adiabatic braiding can be performed in a perfect way, without accidental braiding of unwanted anyons and without thermal excitation of unaccounted anyons. In our proposal, the local operators are expressed in terms of the anyons at the impurity sites, and measurements through the local operators do not involve any mobile or excited anyon. The measurements of the bi-anyonic fusion channel are done through local selective coupling to an extra single channel of chiral current (see Sec. \ref{sec:correlation}), which is non-topological. However, as will be discussed in Sec. \ref{sec:error}, such a measurement is an irrelevant perturbation, which cannot alter the type of anyons the host system supports. Consequently, such a measurement is indeed nondemolitional.

\begin{figure*}[htp!]
	\centering
	\includegraphics[width=\linewidth]{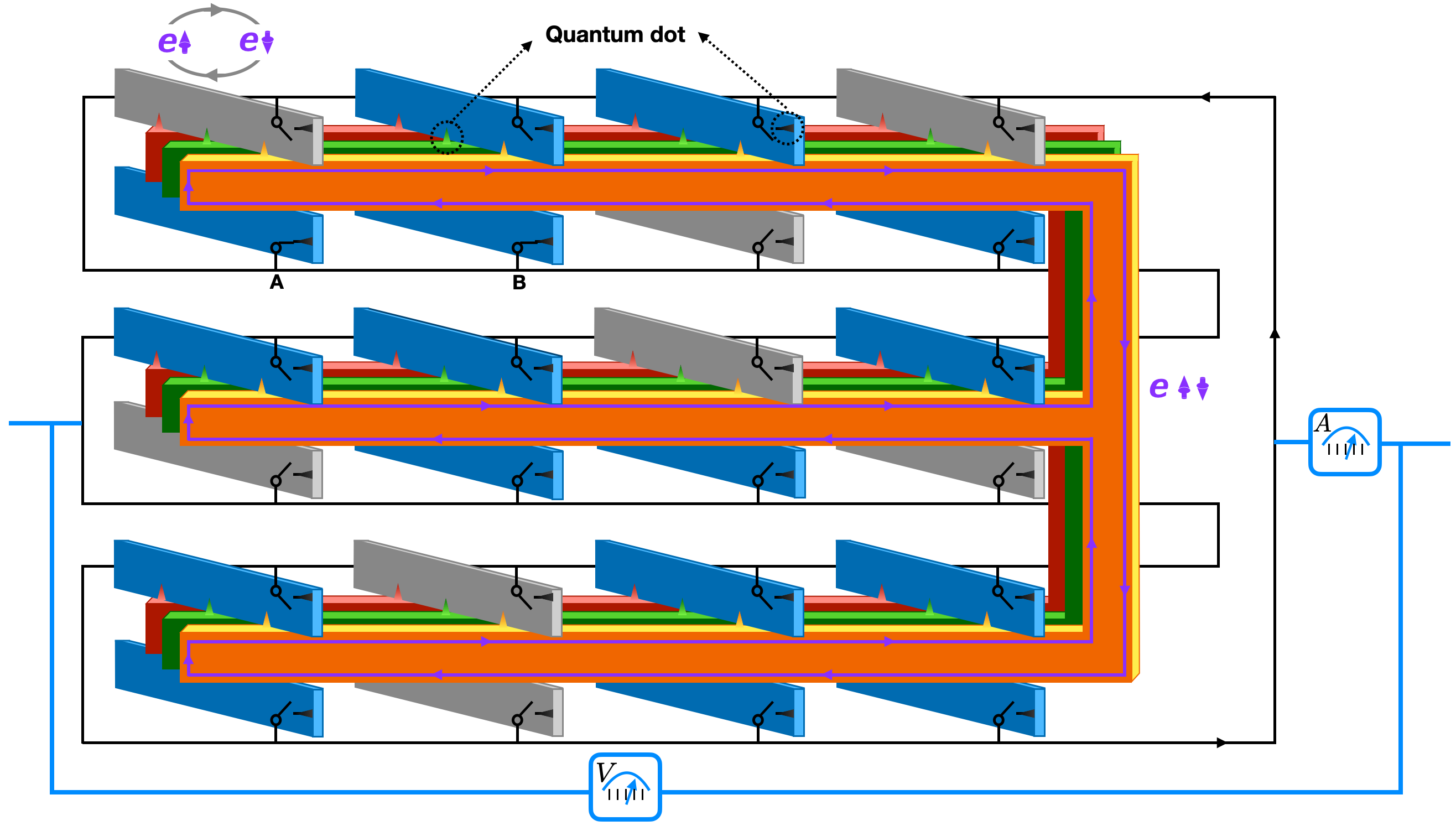}
	\caption{The illustration of the microarchitecture of the quantum computer. The grey and blue slabs are superconducting islands intended for computational and auxiliary units respectively. The orange, green, and red turning-fork slabs are topological insulators. The superconducting islands and the topological insulators are connected by quantum dots, shown as colored wedges. The purple line with arrows depicts the chiral current flowing at the edge of the topological insulator. The black circuit with arrows represents the extra channel of chiral current (the topological insulator  supporting it is not shown for better visualization), whose tunneling to the superconducting islands through quantum dots are controlled by switches. The mechanism of measurement using the black circuit is discussed in Sec. \ref{sec:correlation}. The blue circuit represents the measurement devices. To measure the fusion channel of two anyons localized on the superconducting islands, we just close the corresponding switches in the black circuit, say A and B, and read the voltage and current from the measurement devices. The microarchitecture is designed in this turning-fork shape such that it can be scaled up in a two-dimensional fashion, although it is essentially equivalent to a one-dimensional array shown in Fig. \ref{fig:simple}. The number of grey and blue slabs shown here is just for an illustration of the the scalability. One can extend the structure by having more turning-forks and more grey and blue slabs, arranged in the same fashion as shown here.}
	\label{fig:architecture}
\end{figure*}

The suggested microarchitecture of the quantum computer is schematically shown in Fig. \ref{fig:architecture}. The grey slabs are the superconducting islands that host anyons intended as the computational units, while the blue slabs are the superconducting islands that host anyons intended as the auxiliary units. The Fibonacci anyons are created by coupling to the (pseudo)spin-degenerate chiral modes. There are three chiral modes, and the topological material responsible for each of them are shown in orange, green and red in the figure. The measurements of the bi-anyonic fusion channels are controlled by the switches coupling to yet another chiral mode, shown as the solid black line in the figure, and the measurement apparatus is shown in blue in the figure.
\begin{figure}[htp!]
    \centering
    \includegraphics[width=\linewidth]{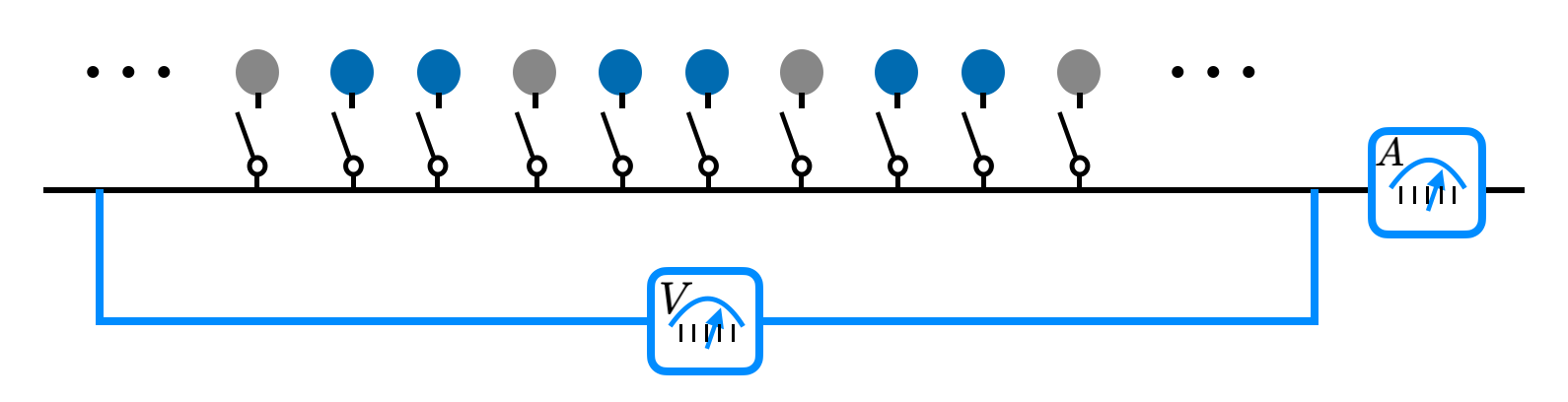}
    \caption{The microarchitecture shown in Fig. \ref{fig:architecture} is equivalent to a one-dimensional array of superconducting islands, shown here as colored dots. The coupling between them and the topological insulators (not shown here for better visualization) creates anyons residing on them. Between every two grey islands (the computational units) there are two blue islands (the auxiliary units). The black line represents the extra channel of chiral current, whose coupling to the superconducting islands is controlled by switches.}
    \label{fig:simple}
\end{figure}

On top of the microarchitecture shown in Fig. \ref{fig:architecture}, the measurement of the fusion channel of two arbitrary anyons can be performed by closing the corresponding switches and decoding the signal. This is possible due to the fact that the signal has features that depend on the fusion channel, as indicated by Eq. (\ref{eq:correlationS}) and (\ref{eq:conductance}). As a result, we need a calibration process performed on every two-unit combination of the microarchitecture components, such that the decoding algorithm for interpreting the signal can be determined. Afterwards, the braiding of two consecutive computational anyons can be performed by the non-deterministic sequence of repeated measurements reviewed in Appendix \ref{app:MBQC}. The braidings of the Fibonacci anyons can then be used to efficiently approximate arbitrary unitary quantum gates \cite{NayakDasSarma,Kitaev_1997,10.5555/2011679.2011685}, as reviewed in Appendix \ref{app:Fibonacci}. The design of the implementation of such approximations up to arbitrarily given precision is called quantum compiling \cite{Maronese2022}, and there is a well-established protocol for Fibonacci anyons \cite{Rouabah_2021}. In conclusion, the quantum computation can be performed on the quantum computer as designed in Fig. \ref{fig:architecture} by programming the switches. Since the switches are simply classical bits, they can be controlled by a classical computer. This consists of a two-layer design of the quantum computer, where the quantum algorithms are executed by manipulating the classical bits, although in a non-deterministic way depending on the quantum output throughout the process.

\subsection{Choice of Materials}

It appears that the main challenge here will be to find a proper material that provides the two-fold degenerate chiral modes. We are aware of two classes of suitable materials, as discussed below.

Firstly, there are topological insulators doped with magnetic ions. The theory suggests Bi$_2$Te$_3$, Bi$_2$Se$_3$, and Sb$_2$Te$_3$ doped with Cr \cite{zhang,science2010}. The low-energy bands of these materials consist of a bonding and an antibonding state of $p_z$ orbitals. Doping with Cr generates ferromagnetism and once the magnetic moments of the Cr ions order, chiral edge states may appear in these topological insulators. It turns out there is a window of parameters when the edge states of the same chirality are not spin polarized. The low energy Hamiltonian of a thin film of such material is given by \cite{science2010} 
\begin{equation}
\begin{split}
  & H = (I + \sigma^z)\Big[\tau^y v k_x + m_+(k)\tau^x + \tau^z vk_y\Big] + \nonumber\\
  & \quad \quad (I - \sigma^z)\Big[-\tau^y v k_x + m_-(k)\tau^x + \tau^z vk_y\Big],  \nonumber\\
  & m_{\pm}(k) = m_0 \pm g_LM - B(k_x^2+k_y^2),
\end{split}
\end{equation}
where $\tau^a$ are Pauli matrices acting on the $p^z$ orbitals $\exp(i\pi/4)\ket{t} \pm \exp(-i\pi/4)\ket{b}$ with $t,b$ standing for the top and the bottom of the layer, $\sigma^z$ acts in the spin space, and $M$ and $g_L$ are the Weiss field and Land\'e factor. At $0< 4(g_LM-m_0)B<1$, there are two edge  modes with identical chirality and opposite spin projections. 

The second class includes the chiral charge-density-wave (CDW) state of kagome metals AV$_3$Sb$_5$ (A= K, Rb, Cs). Some of these materials undergo a phase transition to a 3Q CDW phase at remarkably high temperatures $T_{\text{CDW}} \sim 80-100$ K with complex bond order hosting  loop currents and plaquette fluxes \cite{flux1,flux2}.  The evidence for such a phase comes from the experiments detecting time reversal symmetry breaking without creation of magnetic moments \cite{T-reversal1,T-reversal2,T-reversal3} ascribed to loop currents \cite{Loop1,Loop2}. Such state will have spin degenerate chiral edge modes. In the natural state, however,  the chemical potential lies slightly above the van Hove singularity and as a consequence the CDW state is not insulating. Hence gating will be necessary to enforce the requirements of our model. 

\subsection{Analysis of Errors}
\label{sec:error}

In general, there are two types of errors that should be dealt with carefully in quantum computations: the random error due to coupling to the environment and the unitary error due to imprecision in applying quantum gates. Topological quantum computations using non-Abelian anyons provide hardware-level protection against both types of errors, since the quantum information is stored non-locally among the anyons such that local perturbations cannot alter it \cite{NayakDasSarma}. However, there are still two extra error channels that hinder the actual fault tolerance. One is the wandering of anyons and the other is thermally excited anyonic pairs. They can cause braiding with unwanted anyons and appearance of unaccounted anyons. These difficulties are avoided in the quantum computing platform based on the chiral Kondo chain. As discussed previously, the quantum component of the two-layer design is inherently error-free, since the measurements through local operators do not involve any mobile or excited anyon. On the other hand, the classical component itself can be made fault-tolerant following the well-established protocols for the classical computers. As a result, the remaining error channel lies in the measurement process, where the computing system is coupled to a measurement circuit (the black circuit with switches shown in Fig. \ref{fig:architecture}). For each step of the measurement process, we selectively close two switches and read the signal to interpret the resulting fusion channel of the two selected anyons. In such a step, the computing system is coupled to the measurement circuit locally, as shown in Eq. (\ref{eq:local}). The scaling dimension of the local perturbation $\delta \mathcal{H}_{ij}$ in Eq. (\ref{eq:local}) is larger than 1, thus being irrelevant. This can be understood alternatively via Zamolodchikov's $c$-theorem \cite{Zamolodchikov:1986gt}. According to Eq. (\ref{eq:central}), if we add another chiral edge to a single superconducting island, we will increase the central charge. Here, the chiral edge is added locally to two of the superconducting islands, so the locally enhanced central charge will eventually flow back to the unperturbed value on the global scale. As a result, the effect of the measurement circuit on the computing system is only perturbative, which cannot alter the type of anyons the host system supports, although the associated perturbation itself is non-topological. This leaves us with only the readout error associated with the measurement apparatus (the blue circuit shown in Fig. \ref{fig:architecture}), which can be easily handled by performing projective measurements on the same two anyons repeatedly.

In addition, we have to consider the source of errors associated with the stability of the QCP specific to the Kondo chain system. Firstly, as discussed in Sec. \ref{sec:cft} and Appendix \ref{app:EKpoint}, there is a certain type of Kondo exchange anisotropy that drives the system away from the QCP, which still requires a weaker fine tuning. Secondly, we consider the perturbations lifting the degeneracy between the quantum dots energy levels. This was already discussed in Sec. \ref{sec:model}, but is worth repeating. As we have argued, we should be concerned only with perturbations which couple to the anyons and hence may lift the ground state degeneracy. The most likely perturbation shown in Eq. (\ref{pertD}), which is diagonal in the dot indices, does not even couple to the symplectic spins \bibnotemark[sela]. Moreover, the symplectic symmetry greatly weakens the influence of those perturbations which do couple to the symplectic spins acting as effective magnetic fields. For the $SU(2)$ symmetry such perturbations would present a serious challenge, but for the symplectic symmetry where anyons are weakly coupled to the bulk, they do not. Indeed, according to Eq. (\ref{eq:mutation}) the ``magnetic'' field $H_j$ couples to the anyon zero mode at impurity site $j$ as
\begin{equation}
    \delta \mathcal H = H_j\Gamma_j J_{-1}\Phi_2(x_j).
\end{equation}
This perturbation is strongly irrelevant and as long as one manages to keep it much smaller than the Kondo temperature, its influence will be small \bibnotemark[sela]. The above discussion suggests that the chiral Kondo chain model with symplectic symmetry may indeed present a considerable advantage over previous proposals with unitary symmetry.

Finally, we note that our proposal relies on the assumption that the local operators have a considerable weight in terms of the anyons at the impurity sites, see Eq. (\ref{eq:mutation}). This requires the temperature to be far below the Kondo temperature $T_K\sim E_F\exp\left(-\frac{\pi v_F(E_C - \Delta)}{4(k+1)t^2}\right)$, which may be a major challenge. Besides, the Coulomb interactions between the quantum dots can change the story, so efforts should be taken to mitigate their influences as well.

\section{Summary and Outlook}
\label{sec:summary}

In this paper, we have briefly reviewed the symplectic Kondo effect and extended it to a chiral Kondo chain model. We have showed that under the condition of perfect symplectic symmetry the chiral Kondo chain model is integrable for an arbitrary number of impurities, and possesses non-Abelian anyons in its ground state. At criticality, the local operators can be expressed in terms of the non-Abelian anyons at the impurity sites, such that measurements through the local operators encode the fusion channel of these non-Abelian anyons. We have further analyzed various perturbations around integrability and concluded that except a certain type of Kondo exchange anisotropy analyzed in Appendix \ref{app:EKpoint}, they do not change the low energy behaviors. The symplectic symmetry is responsible for several distinct features of the model, favorable for quantum information applications. It relieves the strict requirement of the $SU(2)$ Kondo effect for the fine tuning of the energy levels and exchange interactions of different channels. Both the anyon contribution to the impurity spin in Eq. (\ref{eq:mutation}) and the coupling between the anyon and the bulk have higher scaling dimensions than those for the unitary symmetry, which drastically reduces the influence of perturbations at criticality. We believe that these features make the chiral Kondo chain with symplectic symmetry a favorable candidate for implementation of the measurement-only topological quantum computing protocol. The essential ingredient of this protocol is the projective measurement of the bi-anyonic fusion channel. We have shown that such measurements can be executed by locally coupling an extra single channel of chiral current to the two anyons and measuring the corresponding conductance at finite frequency. Besides, we have shown that the measurements act as irrelevant perturbations to the host system, which cannot alter the type of anyons the host system supports. We have proposed a two-layer microarchitecture of the quantum computer based on the chiral Kondo chain model with symplectic symmetry, where the quantum algorithms can be executed by manipulating the classical switches in a non-deterministic way. We have also discussed the sources and effects of errors on the proposed quantum computer, and argued that the design is intrinsically fault-tolerant, and the remaining error channels can be handled efficiently. The main challenge to our proposal is to find a proper material that provides the required two-fold degenerate chiral modes, and a moderate Kondo temperature. We have considered two potential candidate materials, with discussion of difficulties associated with their implementations respectively. We look forward to collaborations with experimentalists to push forward the realization of our proposal.

\section{Acknowlegements}

AMT is grateful for D. Goldhaber-Gordon for valuable advice concerning the experimental realization of the suggested model, to A. Weichselbaum, J. von Delft  and to A. Pasupathy and Q. Li for interest in the work. AMT and EJK are grateful for E. Sela and M. Lotem for sharing their preliminary numerical results of the suggested model. AMT and TR are supported by Office of Basic Energy Sciences, Material Sciences and Engineering Division, U.S. Department of Energy (DOE) under Contract No. DE-SC0012704.

\appendix

\section{Emery-Kivelson style analysis of the anisotropic $k=2$ case}
\label{app:EKpoint}

We address the problem of  stability of the $Sp(2k)$ QCP for the case of $k=2$ using the Emery-Kivelson approach \cite{EK1992}. Namely, we will consider the case of large anisotropy with large coupling constants for the Kac-Moody currents belonging to the Cartan subalgebra and perform bosonization and refermionization. At certain special value of the Cartan couplings, the Hamiltonian becomes quadratic in the fermionic operators, thus allowing a complete analysis.
 
The generators of the $\mathfrak{sp}(4)$ algebra can be represented as tensor products of Pauli matrices: 
\begin{equation}
    (I\otimes\sigma^a), ~~(\tau^z\otimes\sigma^a), ~~ (\tau^x\otimes\sigma^a), ~~ (\tau^y\otimes I),
\end{equation}
where $a=x,y,z$. For future conveniences we will interchange $\tau^z$ and $\tau^y$. Then the generators become $(I\otimes\sigma^a), ~(\tau^y\otimes\sigma^a), ~(\tau^x\otimes\sigma^a), ~ (\tau^z\otimes I)$. We further bosonize the fermions:
\begin{equation}
\begin{split}
    & c_{i\sigma} = \frac{\kappa_{i\sigma}}{\sqrt{2\pi a_0}}\exp\Big[- \ri\sqrt{\pi}(\varphi_c +i\varphi_{f} + \sigma\varphi_s +i\sigma\varphi_{sf})\Big],\\
    &  i,\sigma = \pm 1, \quad [\varphi(x),\varphi(y)] = -\frac{\ri}{4}\operatorname{sgn}(x-y),
\end{split}
\end{equation}
where we assign $+,-$ to $\uparrow,\downarrow$ respectively, $\varphi$ in the commutator stands for any of $\varphi_c,\varphi_f,\varphi_{sf}$, and $\{\kappa_{i\sigma},\kappa_{i'\sigma'}\} = 2\delta_{ii'}\delta_{\sigma\sigma'}$ are Klein factors.

The Klein factors are 4-dimensional Dirac gamma matrices. Their irreducible representation requires the following restriction 
\begin{equation}
\kappa_{+\uparrow}\kappa_{+\downarrow}\kappa_{-\uparrow}\kappa_{-\downarrow} =1.
\end{equation}
Then we can construct all 10 Kac-Moody currents: 
\begin{equation}
\label{eq:algebra}
\begin{split}
& I\otimes\sigma^+ = c^+_{i\uparrow}c_{i\downarrow} = 2\ri \kappa_{+\uparrow}\kappa_{+\downarrow}\re^{\ri\sqrt{4\pi}\varphi_s}\sin(\sqrt{4\pi}\varphi_{sf}), \\
& I\otimes\sigma^- = c^+_{i\downarrow}c_{i\uparrow} = 2\ri \kappa_{+\uparrow}\kappa_{+\downarrow}\re^{-\ri\sqrt{4\pi}\varphi_s}\sin(\sqrt{4\pi}\varphi_{sf}), \\
& \tau^+\otimes\sigma^z = \frac{1}{2}(c^+_{+\uparrow}c_{-\uparrow} - c^+_{+\downarrow}c_{-\downarrow}) \\
& \quad \quad \quad \quad = \ri\kappa_{+\uparrow}\kappa_{-\uparrow}\re^{\ri\sqrt{4\pi}\varphi_f}\sin(\sqrt{4\pi}\varphi_{sf}), \\
& \tau^-\otimes\sigma^z = \frac{1}{2}(c^+_{-\uparrow}c_{+\uparrow} - c^+_{-\downarrow}c_{+\downarrow}) \\
& \quad \quad \quad \quad = \ri\kappa_{+\uparrow}\kappa_{-\uparrow}\re^{-\ri\sqrt{4\pi}\varphi_f}\sin(\sqrt{4\pi}\varphi_{sf}), \\
& \tau^+\otimes\sigma^+ = \kappa_{+\uparrow}\kappa_{-\downarrow}\re^{\ri\sqrt{4\pi}(\varphi_f +\varphi_s)},\\
& \tau^+\otimes\sigma^- = \kappa_{+\uparrow}\kappa_{-\downarrow}\re^{\ri\sqrt{4\pi}(\varphi_f -\varphi_s)},\\
& \tau^-\otimes\sigma^+ = -\kappa_{+\uparrow}\kappa_{-\downarrow}\re^{\ri\sqrt{4\pi}(-\varphi_f +\varphi_s)},\\
& \tau^-\otimes\sigma^- =- \kappa_{+\uparrow}\kappa_{-\downarrow}\re^{-\ri\sqrt{4\pi}(\varphi_f +\varphi_s)},\\
& I\otimes\sigma^z \sim \partial_x\varphi_s, ~~ \tau^z\otimes I \sim \partial_x\varphi_f. 
\end{split}
\end{equation}
We introduce one Majorana and two Dirac fermions:
\begin{eqnarray}
&& \xi = \frac{\kappa_{+\uparrow}}{\sqrt{2\pi a_0}}\sin(\sqrt{4\pi}\varphi_{sf}), \nonumber\\
&& \psi_s = \frac{\kappa_{+\downarrow}}{\sqrt{2\pi a_0}}\re^{\ri\sqrt{4\pi}\varphi_s}, ~~ \psi_f = \frac{\kappa_{-\uparrow}}{\sqrt{2\pi a_0}}\re^{\ri\sqrt{4\pi}\varphi_f},
\end{eqnarray}
which corresponds to five Majorana fermions. Consequently, the currents in Eq. (\ref{eq:algebra}) are expressed in terms of these five Majoranas.

Below we separate all possible interactions into two terms. The term $V_{\perp}$ contains non-Cartan generators and all couplings there are small; the term $V_{\parallel}$ contains Cartan generators and, in accordance with the Emery-Kivelson approach, their couplings are large. For the sake of brevity we omit coordinate  in the expression for $\varphi$ fields assuming that in the expressions below they are positioned at the the impurity site $x=0$. Using the bosonization rules in Eq. (\ref{eq:algebra}) we arrive at:
\begin{widetext}
    \begin{equation}
    \label{eq:allinteractions}
    \begin{split}
    V_{\perp} = &\frac{\ri J_s }{\sqrt{2\pi a_0}}\xi(0)\Big[ \kappa_{+\downarrow}\re^{\ri\sqrt{4\pi}\varphi_s}(I\otimes\sigma^-) + \kappa_{+\downarrow}\re^{-\ri\sqrt{4\pi}\varphi_s}(I\otimes\sigma^+)\Big] \\
    + &\frac{\ri J_f}{\sqrt{2\pi a_0}}\xi(0)\Big[ \kappa_{-\uparrow}\re^{\ri\sqrt{4\pi}\varphi_f}(\tau^-\otimes\sigma^z) + \kappa_{-\uparrow}\re^{-\ri\sqrt{4\pi}\varphi_f}(\tau^+\otimes\sigma^z)\Big] \\
    + & \frac{\kappa_{+\uparrow}\kappa_{-\downarrow}}{2\pi a_0}\Big\{ (J' +J'')\Big[\re^{\ri\sqrt{4\pi}(\varphi_f +\varphi_s)} (\tau^-\otimes\sigma^-) + \re^{-\ri\sqrt{4\pi}(\varphi_f +\varphi_s)} (\tau^+\otimes\sigma^+)\Big] \\
    & \quad \quad \quad +(J' -J'')\Big[\re^{\ri\sqrt{4\pi}(-\varphi_f +\varphi_s)} (\tau^+\otimes\sigma^-) + \re^{\ri\sqrt{4\pi}(\varphi_f -\varphi_s)} (\tau^-\otimes\sigma^+)\Big]\Big\},\\
    V_{\parallel} =& \frac{J_{\parallel}}{\sqrt\pi}\Big[\partial_x\varphi_s (I\otimes\sigma^z) + \partial_x\varphi_f (\tau^z\otimes I)\Big].
    \end{split}
    \end{equation}
\end{widetext}
Now we absorb the complex exponents into the sigma- and tau-matrices:
\begin{equation}
 \sigma^{\pm}\re^{\mp\ri\sqrt{4\pi}\varphi_s} = \tilde\sigma^{\pm}, ~~ \tau^{\pm}\re^{\mp\ri\sqrt{4\pi}\varphi_f} = \tilde\tau^{\pm},
\end{equation}
which leads to a shift of $J_{\parallel}$ and Eq. (\ref{eq:allinteractions}) becomes:
\begin{equation}
\label{eq:V}
\begin{split}
 V_{\perp} = & 2\ri J_s \xi(0)\kappa_{+\downarrow}(I\otimes\tilde{\sigma}^x) + 2\ri J_f \xi(0)\kappa_{-\uparrow}(\tilde{\tau}^x\otimes\tilde{\sigma}^z)\\
 + & 2 \kappa_{+\downarrow}\kappa_{-\uparrow}\Big[J'(\tilde{\tau}^x\otimes\tilde{\sigma}^x) + J''(\tilde{\tau}^y\otimes\tilde{\sigma}^y)\Big],\\
 V_{\parallel} = &\frac{(J_{\parallel}- \pi v_F)}{\sqrt\pi}\Big[\partial_x\varphi_s (I\otimes\tilde{\sigma}^z) + \partial_x\varphi_f(\tilde{\tau}^z\otimes I)\Big],
 \end{split}
\end{equation}
where $J_s,J_f,J'$ and $J''$ are rescaled such that the resulting expressions are compact. We then define the new local fermions as 
\begin{equation}
\label{eq:local}
\begin{split}
&\rho_1 = \kappa_{+\downarrow}(I\otimes\tilde\sigma^x), \quad \rho_2= \kappa_{-\uparrow}(\tilde\tau^x\otimes I),\\
&\eta_1 = \kappa_{+\downarrow}(I\otimes\tilde\sigma^y), \quad \eta_2 =\kappa_{-\uparrow}(\tilde\tau^y\otimes I),
\end{split}
\end{equation}
so that they anticommute with each other and with the itinerant fermion $\xi$. Then $V_{\perp}$ and $V_{\parallel}$ can be rewritten as
\begin{eqnarray}
    && V_{\perp}= 2\ri J_s\xi(0)\rho_1 + 2 J'\rho_1\rho_2 + 2 J''\eta_1\eta_2 \nonumber\\
    && \quad \quad ~-2 J_f\xi(0)\rho_1\rho_2\eta_1, \label{eq:rel} \\
    && V_{\parallel}= \frac{\ri(J_{\parallel}- \pi v_F)}{\sqrt\pi}[\partial_x\varphi_s\eta_1\rho_1 + \partial_x\varphi_f\eta_2\rho_2]. \label{eq:irr}
\end{eqnarray}
The four-fermion interaction $\xi(0)\rho_1\rho_2\eta_1$ in Eq. (\ref{eq:rel}) can be reduced by replacing $\rho_1\rho_2$ with its average, then the effective action at the Emery-Kivelson point $J_\parallel=\pi v_F$ becomes quadratic in fermions:
\begin{equation}
\begin{split}
& S = \Psi^T\left(
\begin{array}{ccccc}
G^{-1}(\omega) & \ri J_s & \ri J_f & 0 & 0\\
-\ri J_s &\omega &0 & J' &0\\
 -\ri J_f & 0 &\omega &0 &  J''\\
0 & - J' & 0 &\omega &0\\
0 & 0 &- J'' &0 &\omega
\end{array}
\right)\Psi, \\
& \Psi^T =(\xi(0), \rho_1,\eta_1,\rho_2,\eta_2) 
~~G(\omega) = \ri\pi\nu(\epsilon_F)\operatorname{sgn}\omega,
\end{split}
\end{equation}
It can be seen that the simultaneous presence of $J'$ and $J''$ are crucial for the instability of the QCP, while with one of these terms being absent the ground state is critical. For example, at $J''=0$, the Majorana fermion $\eta_2$ decouples, which gives us the ground state entropy $S(0) = \ln\sqrt 2$. Then at small frequencies we have the following expression for the Green's function of the remaining fermions $(\xi,\rho_1,\eta_1,\rho_2)$:
\begin{widetext}
\begin{equation}
{\cal G} =\frac{1}{-\ri\pi\nu{J_f}^2\operatorname{sgn}\omega+ \omega}\begin{pmatrix}
 		\ri\pi\nu|\omega| & 0 & \pi\nu  {J_f}\operatorname{sgn}\omega & -\pi\nu{J_s}|\omega|/{J'}  \\
  		0 & -\ri\pi\nu {J_f}^2 |\omega|/{J'}^2  & \ri\pi\nu {J_f} {J_s} |\omega|/{J'}^2  & \ri\pi\nu{J_f}^2/{J'} \operatorname{sgn}\omega- \omega/{J'}  \\
 		-\pi\nu{J_f}\operatorname{sgn}\omega & \ri\pi\nu {J_f} {J_s} |\omega|/{J'}^2  & 1-\ri\pi\nu {J_s}^2 |\omega|/{J'}^2 & -\ri\pi\nu {J_f} {J_s}/{J'}\operatorname{sgn}\omega \\
 		-\pi\nu {J_s} |\omega| /{J'} & -\ri\pi\nu  {J_f}^2/{J'}\operatorname{sgn}\omega+ \omega/{J'}  & \ri\pi\nu {J_f} {J_s}/{J'}\operatorname{sgn}\omega & -\ri\pi\nu |\omega| ({J_f}^2+{J_s}^2)/{J'}^2 \end{pmatrix},
\end{equation}
\end{widetext}
The corrections to the free energy from the irrelevant operators are given by (\ref{eq:irr}). The asymptotic behaviors of the relevant components of $\mathcal{G}$ at large $\tau$ are
\begin{equation}
\begin{split}
&{\cal G}_{\eta_1\eta_1} (\tau) \sim 1/\tau, ~~{\cal G}_{\rho_1\rho_1}(\tau) \sim 1/\tau^2 ,\\
& {\cal G}_{\eta_2,\eta_2}(\tau) \sim \operatorname{sgn}\tau, ~~{\cal G}_{\rho_2\rho_2}(\tau)\sim 1/\tau^2
\end{split}
\end{equation}
then the first term of Eq. (\ref{eq:irr}) does not yield any singularities. For the second term of Eq. (\ref{eq:irr}), although ${\cal G}_{\eta_2\eta_2}(\tau) \sim \operatorname{sgn}\tau$, it is compensated by a very fast decay of ${\cal G}_{\rho_2\rho_2}(\tau)\sim 1/\tau^{2}$. As a result, for $J''=0$ the deviations from the Emery-Kivelson point does not generate any strong non-analytic terms in the thermodynamics, thus the symplectic QCP exists for arbitrary ratios of $J_s, J_f, J'$. 

Let us now consider both $J',J'' \neq 0$, where the Green's function couples all five fermions $(\xi,\rho_1,\eta_1,\rho_2,\eta_2)$, and at small frequencies it has the following expression:
\begin{widetext}
\begin{equation}
\begin{split}
&{\cal G}=\frac{1}{{J'}^2{J''}^2-\ri\pi\nu({J'}^2{J_f}^2+{J''}^2{J_s}^2)|\omega|} \\
        &\begin{pmatrix}
		\ri\pi\nu {J'}^2 {J''}^2\operatorname{sgn}\omega & \pi\nu {J''}^2 {J_s} |\omega|  & \pi\nu {J'}^2 {J_f} |\omega| & -\pi\nu {J'} {J''}^2 {J_s}\operatorname{sgn}\omega  & -\pi\nu {J'}^2 {J''} {J_f}\operatorname{sgn}\omega \\
 		-\pi\nu {J''}^2 {J_s} |\omega|  & {J''}^2 \omega  & 0 & -{J'}{J''}^2+\ri\pi\nu {J'} {J_f}^2 |\omega|  & -\ri\pi\nu {J''} {J_f} {J_s} |\omega| \\
 		-\pi\nu {J'}^2 {J_f} |\omega| & 0 & {J'}^2 \omega & -\ri\pi\nu {J'}{J_f} {J_s} |\omega|  & -{J'}^2 {J''}+\ri\pi\nu {J''} {J_s}^2 |\omega| \\
 		-\pi\nu {J'} {J''}^2 {J_s}\operatorname{sgn}\omega & {J'} {J''}^2-\ri\pi\nu {J'} {J_f}^2 |\omega| & \ri\pi\nu{J'} {J_f} {J_s} |\omega| & -\ri\pi\nu {J''}^2 {J_s}^2 \operatorname{sgn}\omega +{J''}^2 \omega  & -\ri\pi\nu {J'} {J''} {J_f} {J_s}\operatorname{sgn}\omega \\
 		-\pi\nu {J'}^2 {J''} {J_f}\operatorname{sgn}\omega & \ri\pi\nu {J''} {J_f} {J_s} |\omega| & {J'}^2{J''}-\ri\pi\nu {J''} {J_s}^2 |\omega|  & -\ri\pi\nu {J'} {J''} {J_f} {J_s}\operatorname{sgn}\omega & -\ri\pi\nu {J'}^2{J_f}^2 \operatorname{sgn}\omega +{J'}^2 \omega 
	\end{pmatrix}.
\end{split}
\end{equation}
\end{widetext}
The asymptotic behaviors of the Green's functions for all local fermions at large $\tau$  become Fermi liquid like. The corresponding energy scale is $1/(\pi\nu {J_f}^2/{J''}^2+\pi\nu {J_s}^2/{J'}^2)$, where $\pi\nu{J_f}^2$ and $\pi\nu{J_s}^2$ play the role of the Kondo temperatures for the anisotropic couplings.

Our calculation for the $Sp(4)$ symmetry presented above demonstrates that, in the limit of strong anisotropy of couplings of the Cartan generators, there is a single direction in the generator parameter space along which the model scales away from the symplectic QCP and to a Fermi liquid. 

\section{Correlation of $Sp_1(2k)$ Primary Fields}
\label{app:primarycorrelation}

The singular behavior of the correlation functions of $Sp_1(2k)$ primary fields at short distance can be determined by a powerful method called the boundary conformal field theory (boundary CFT) \cite{CARDY1984514,CARDY1989581,CARDY1991274}. This method expresses the effect of the impurity spins as conformally invariant boundary conditions on the chiral fermions, and further trades the boundary CFT for a bulk CFT with boundary states. There is a bijection between the primary fields in the bulk CFT and the conformally invariant boundary conditions, such that boundary states can be constructed by fusing the bulk primary fields (here the spin of the chiral fermions) with a reference boundary state (here the impurity spin). This has been used to tackle the problem of a single impurity spin, known as the fusion ansatz \cite{AFFLECK1990517,AFFLECK1991849,AFFLECK1991641,PhysRevLett.68.1046}, which we now review.

We here denote the boundary condition using upper case $A$, and the conformal tower of the primary field $\Phi_a$ using lower case $a$. The boundary state $\ket{A}$ is then expanded in terms of the so-called Ishibashi state $\ket{a}$ constructed from the conformal tower $a$ \cite{doi:10.1142/S0217732389000320}, and the expansion coefficients $\langle\langle a|A\rangle$ fulfill the following Cardy conditions \cite{https://doi.org/10.48550/arxiv.hep-th/0411189}:
\begin{equation}
    \sum_b S^a_b n^b_{AB}=\langle A| a\rangle\rangle \langle\langle a|B\rangle,
\end{equation}
where $S^a_b$ is the modular $S$-matrix of the bulk CFT \cite{CFT}, and $n^b_{AB}$ is the multiplicity of the conformal tower $b$ under the boundary conditions $A$ and $B$. Utilizing the Verlinde formula for the modular $S$-matrix \cite{VERLINDE1988360}, the Cardy conditions can be solved by the following fusion ansatz
\begin{equation}
\label{eq:fusionansatz}
    \langle\langle a | B\rangle=\langle\langle a | A\rangle \frac{S_c^a}{S_0^a}, \quad n^a_{AB}=\sum_bN_{bc}^{a}n^b_{AA},
\end{equation}
where the boundary state $\ket{B}$ is obtained by fusing the boundary state $\ket{A}$ with the primary field $\Phi_c$, $0$ denotes the vacuum, and the coefficients $N^a_{bc}$ are those appearing in the fusion rule for the bulk primary fields:
\begin{equation}
    \Phi_a \times \Phi_b=\sum_c N_{ab}^c \Phi_c.
\end{equation}
Using the fusion ansatz in Eq. (\ref{eq:fusionansatz}), the correlation between the primary fields can be evaluated \cite{CARDY1991274}:
\begin{equation}
\label{eq:correlator1}
\begin{split}
    \left\langle\Phi_a\left(z_1\right) \Phi_a\left(z_2\right)\right\rangle&=\frac{1}{\left(z_1-z_2\right)^{2 h_a}} \frac{\langle\langle a | B\rangle}{\langle\langle 0 | B\rangle}\\
    &=\frac{1}{\left(z_1-z_2\right)^{2 h_a}}\frac{S^a_c/S^a_0}{S^0_c/S^0_0},
\end{split}
\end{equation}
where $h_a$ is the scaling dimension of the primary field $\Phi_a$, the impurity spin is placed at the origin, the complex coordinate is defined as $z\equiv \tau-\ri x$, and the correlation is evaluated at $\operatorname{Im}z_1<0$ and $\operatorname{Im}z_2>0$ around the impurity spin. Eq. (\ref{eq:correlator1}) can be easily evaluated as long as the modular $S$-matrix for the primary fields is known. 

The above described fusion ansatz can be readily applied to the case of a single impurity spin. For the case of multiple impurity spins, a generalized multi-fusion ansatz is proposed under the condition that each impurity spin affects the chiral fermions independently \cite{Sela1,Sela2}:
\begin{equation}
\label{eq:multi}
    n_{F M}^a=\sum_b N_{b c}^a n_{F M-1}^b,
\end{equation}
where $F$ refers to the free boundary condition on the left, $M$ refers to the boundary condition of $M$ impurity spins on the right, and $M$ is obtained by fusing $(M-1)$ with the primary field $\Phi_c$. Iterating Eq. (\ref{eq:multi}), and using the analogy with the fusion category theory, an effective primary field $\Phi_{c_{\text{eff}}}$ and an effective boundary $B_{\text{eff}}$ can be defined to represent the collective effect of the $M$ impurity spins:
\begin{equation}
\begin{split}
    & n_{F M}^a = \sum_{c_{\text {eff}}} \operatorname{dim}\left[V_{c \ldots c}^{c_{\text {eff}}}\right] n_{F B_{\text {eff}}}^a, \\
    & \underbrace{c \times c \times \ldots \times c}_M=\sum_{c_{\text{eff}}}\operatorname{dim}\left[V_{c \ldots c}^{c_{\mathrm{eff}}}\right] c_{\mathrm{eff}}, \\
    & n_{F B_{\mathrm{eff}}}^a=\sum_bN_{b c_{\mathrm{eff}}}^a n_{F F}^b,
\end{split}
\end{equation}
where $\operatorname{dim}\left[V_{c \ldots c}^{c_{\text {eff}}}\right]$ is the dimension of the Hilbert space obtained by $(M-1)$ fusions of $c$ into $c_{\text{eff}}$. Consequently, the correlation between the primary fields $\Phi_a$ can be evaluated as
\begin{equation}
\label{eq:phiphi}
\begin{split}
    \left\langle\Phi_a\left(z_1\right) \Phi_a\left(z_2\right)\right\rangle &=\frac{1}{\left(z_1-z_2\right)^{2 h_a}} \frac{\left\langle\langle a | B_{\mathrm{eff}}\right\rangle}{\left\langle\langle 0 | B_{\mathrm{eff}}\right\rangle}\\
    &=\frac{1}{\left(z_1-z_2\right)^{2 h_a}} \frac{S_{c_{\mathrm{eff}}}^a / S_0^a}{S_{c_{\mathrm{eff}}}^0 / S_0^0},
\end{split}
\end{equation}
where $z_1$ sits to the left of the $M$ impurity spins, and $z_2$ sits to the right of the $M$ impurity spins. In case of the $Sp_1(2k)$ theory, the modular-$S$ matrix is \cite{KAC1988156}
\begin{equation}
    S_{a}^{a^{\prime}}=\sqrt{\frac{2}{2+k}} \sin \left[\frac{\pi(a+1)\left(a^{\prime}+1\right)}{2+k}\right].
\end{equation}
As a result, the correlation function in Eq. (\ref{eq:phiphi}) encodes the information of the total fusion channel of $M$ anyons localized at the impurity sites, through the correspondence between the primaries and the anyons. Such a correspondence can be checked by looking at the fusion rules for the $Sp_1(2k)$ primaries shown in Eq. (\ref{eq:fusionrule}), and the results for the two relevant cases $k=2$ and $k=3$ are shown in Eq. (\ref{eq:k2Ising}) and (\ref{eq:k3Fibonacci}) respectively.

\section{Fibonacci Anyons}
\label{app:Fibonacci}

\begin{figure}[htp!]
	\centering
	\includegraphics[width=\linewidth]{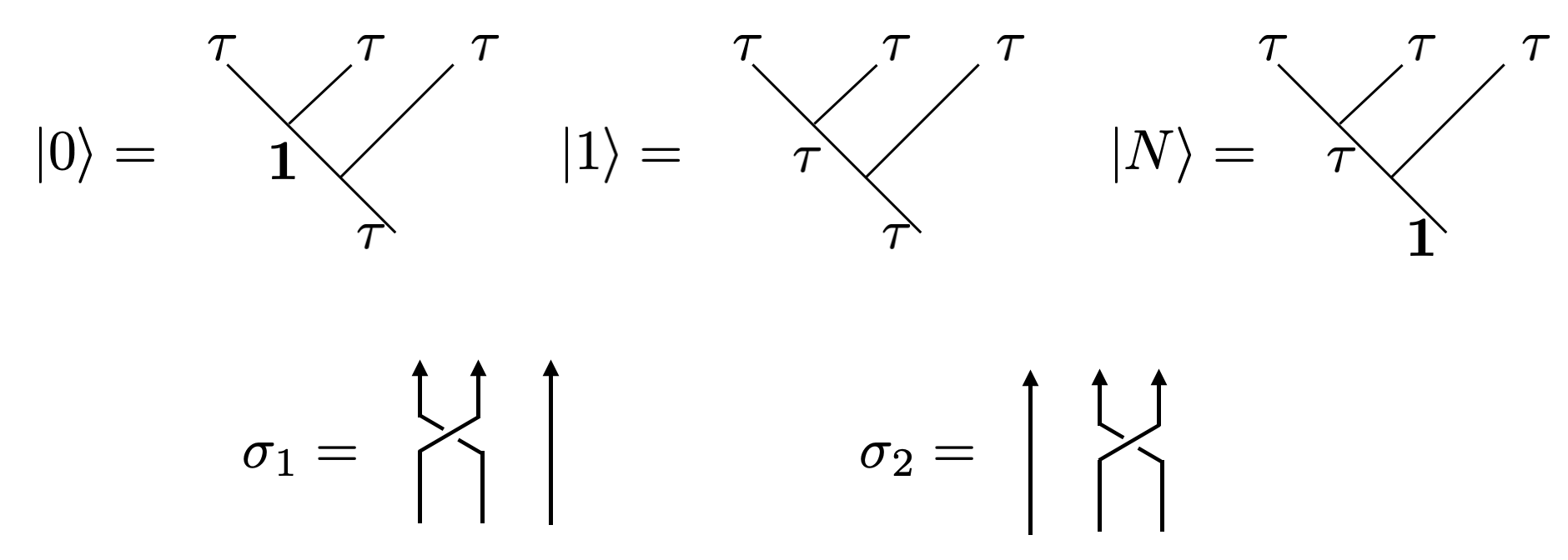}
	\caption{The computational basis and the braiding operators using three anyons. The states $\ket{0}$ and $\ket{1}$ constitute two levels of the qubit, while the state $\ket{N}$ is a non-computational auxiliary state.}
	\label{fig:basis}
\end{figure}

In addition to the vacuum $\mathbf{1}$, there is a single type of anyon $\tau$. The fusion rules are
\begin{equation}
    \tau\times \mathbf{1}=\mathbf{1}\times \tau=\tau, \quad \tau \times \tau=\mathbf{1}+\tau.
\end{equation}
The quantum dimension is
\begin{equation}
    d_{\tau}=\frac{1+\sqrt{5}}{2}.
\end{equation}
The non-Abelian nature of the Fibonacci anyons is contained in its nontrivial fusion matrix:
\begin{equation}
    F_{\tau}^{\tau\tau\tau}=\begin{pmatrix}
    d^{-1}_{\tau} & d^{-1/2}_{\tau} \\ d^{-1/2}_{\tau} & -d^{-1}_{\tau}
    \end{pmatrix}.
\end{equation}
The nontrivial braiding matrices for the Fibonacci anyons are
\begin{equation}
    R_{\mathbf{1}}^{\tau\tau}=e^{\ri 4\pi/5}, \quad R_{\tau}^{\tau\tau}=e^{-\ri 3\pi /5}.
\end{equation}
For implementation of two-level gates, we need three anyons \cite{NayakDasSarma}. The three anyon basis can be chosen as shown in the upper panel of Fig. \ref{fig:basis}, and the braiding operators are shown in the lower panel of Fig. \ref{fig:basis}.
The two braiding operators can be represented in terms of the three anyons basis as \cite{NayakDasSarma}
\begin{equation}
\begin{split}
    & \rho\left(\sigma_1\right)\begin{pmatrix}
|0\rangle \\
|1\rangle \\
|N\rangle
\end{pmatrix}=\begin{pmatrix}
e^{-4 \pi i / 5} & 0 & 0 \\
0 & e^{3 \pi i / 5} & 0 \\
0 & 0 & e^{3 \pi i / 5}
\end{pmatrix}\begin{pmatrix}
|0\rangle \\
|1\rangle \\
|N\rangle
\end{pmatrix}, \\
    & \rho\left(\sigma_2\right)\begin{pmatrix}
|0\rangle \\
|1\rangle \\
|N\rangle
\end{pmatrix}=\begin{pmatrix}
-\frac{e^{-\pi i / 5}}{d_{\tau}} & -\frac{i e^{-i \pi / 10}}{\sqrt{d_{\tau}}} & 0 \\
-\frac{i e^{-i \pi / 10}}{\sqrt{d_{\tau}}} & -\frac{1}{d_{\tau}} & 0 \\
0 & 0 & e^{3 \pi i / 5}
\end{pmatrix}\begin{pmatrix}
|0\rangle \\
|1\rangle \\
|N\rangle
\end{pmatrix}.
\end{split}
\end{equation}
It is easy to see that the states $\ket{0}$ and $\ket{1}$ constitute the computational basis, while the state $\ket{N}$ stands alone as an non-computational auxiliary state. According to the Solovay-Kitaev theorem \cite{Kitaev_1997,10.5555/2011679.2011685}, any unitary quantum gate can be approximated by the two braiding operators up to arbitrary precision efficiently. For example, the Hadamard gate can be approximated as \cite{Rouabah_2021}:
\begin{equation}
\begin{split}
    W_H & =\sigma_1^{+4} \sigma_2^{-2} \sigma_1^{+2} \sigma_2^{-2} \sigma_1^{+2} \sigma_2^{+2} \sigma_1^{-2} \sigma_2^{+4} \sigma_1^{+2} \sigma_2^{-2} \sigma_1^{-2} \sigma_2^{+2} \sigma_1^{+2} \\
    & =-\frac{i}{\sqrt{2}}\left(\begin{array}{cc}
1.0040+0.0056 i & 0.9959-0.0048 i \\
0.9959+0.0048 i & -1.0040+0.0056 i
\end{array}\right),
\end{split}
\end{equation}
where the extra phase is not a problem, since eventually a projective measurement will be performed.

\section{Measurement-Only Topological Quantum Computation}
\label{app:MBQC}

 We review the basic ingredients of the measurement-only topological quantum computation proposed in Ref. \onlinecite{BondersonNayak2008} in terms of Fibonacci anyons, adopting the diagrammatic notations used by Kitaev \cite{KITAEV20062,BondersonNayak2008,BONDERSON20082709}. It includes the main components listed below.

\textit{Orthogonal projective measurement of fusion channel} $c\in \{\mathbf{1},\tau\}$ \textit{of a state} $\ket{\psi}$:
\begin{equation*}
\begin{split}
	& \operatorname{Prob}(c)=\left\langle\psi\left|\Pi_c\right| \psi\right\rangle, \quad |\psi\rangle \mapsto \frac{\Pi_c|\psi\rangle}{\sqrt{\left\langle\psi\left|\Pi_c\right| \psi\right\rangle}}, \\
    & \Pi_a\Pi_b=\delta_{ab}\Pi_a,
\end{split}
\end{equation*}
whose diagrammatic representation is
\begin{center}
    \includegraphics[scale=0.35]{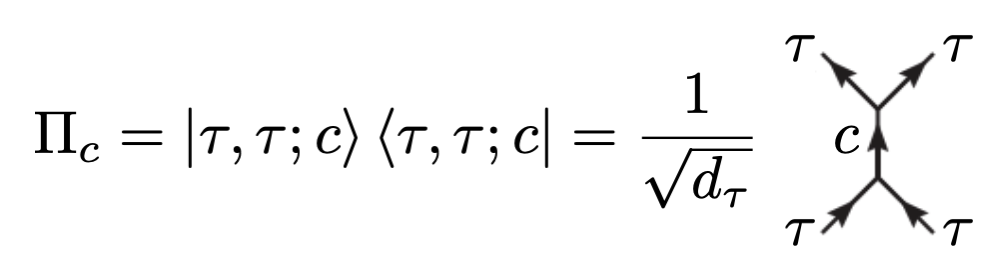}
\end{center}
Conventionally, the line for the vacuum $c=\mathbf{1}$ is omitted for better visualization. In our proposal, the set of orthogonal projective measurements is realized by measuring the bi-anyonic fusion channel through the correlation function of the impurity spins, see Sec. \ref{sec:correlation}.

\textit{The information-encoded state} $\psi(\tau,\cdots)$
\begin{center}
    \includegraphics[scale=0.35]{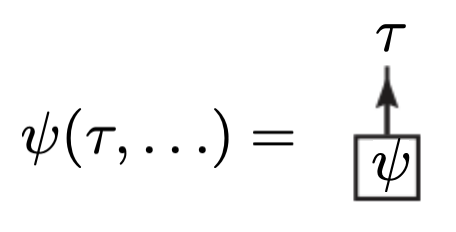}
\end{center}
where only the anyon on which we want to perform the operation is shown.

\textit{Maximally entangled Bell state} $\ket{\tau,\tau;\mathbf{1}}$
\begin{center}
    \includegraphics[scale=0.35]{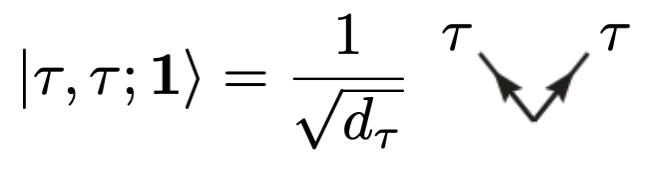}
\end{center}
where the line for the vacuum is conventionally omitted.

Firstly, let us consider the quantum teleportation using three anyons, which is diagrammatically represented as
\begin{center}
    \includegraphics[scale=0.35]{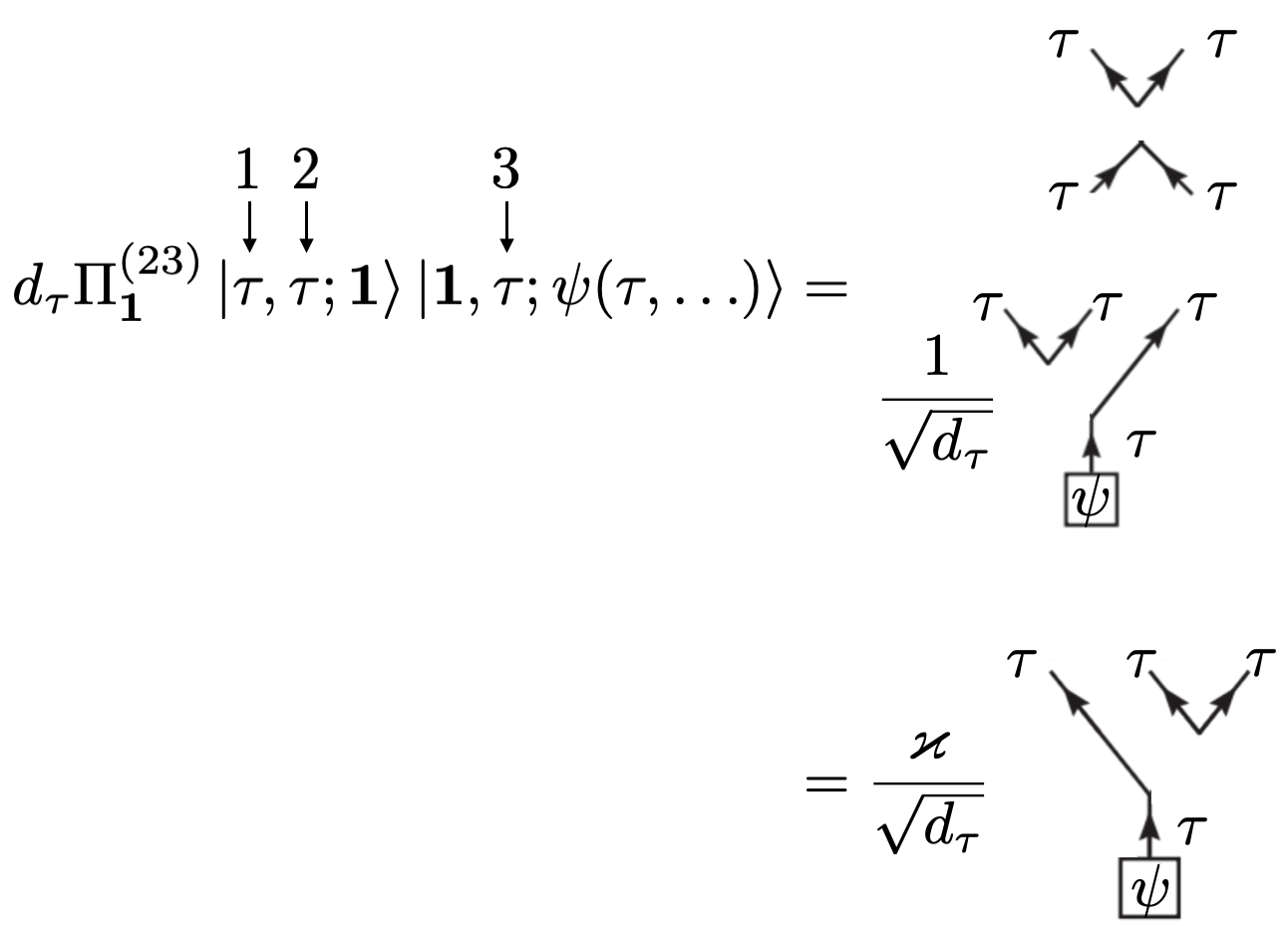}
\end{center}
where $\varkappa$ is a pure phase. In this way, the information-encoded in anyon 3 with others of $\psi$ is transported to anyon 1 with others of $\psi$, using anyon 2 as an auxiliary anyon. However, the measurement of anyons 2 and 3 may not be in the vacuum channel, this is where the repeated measurements come in. We denote the fusion outcome of anyons 1 and 2 as $e_i$, and the one of anyons 2 and 3 as $f_i$. Imagine we start with $e_1=\mathbf{1}$, and we measure anyons 2 and 3 with outcome $f_1$. If $f_1=\mathbf{1}$, then we are done; otherwise we measure anyons 1 and 2 with outcome $e_2$. This gives us a sequence of outcomes $(e_1,f_1;e_2,f_2;\ldots)$. The average number of attempts until $f_n=\mathbf{1}$ is $\langle n\rangle \leqslant d^2_{\tau}<3$, and the probability of needing $n>N$ attempts is
\begin{equation*}
    \operatorname{Prob}\left(f_1, \ldots, f_N \neq \bm{1}\right) \leq \left(1-d_{\tau}^{-2}\right)^N \approx 0.618^N
\end{equation*}
which is exponentially suppressed. Such a non-deterministic operation is dubbed as a forced measurement, and it is diagrammatically represented as
\begin{center}
	\includegraphics[scale=0.35]{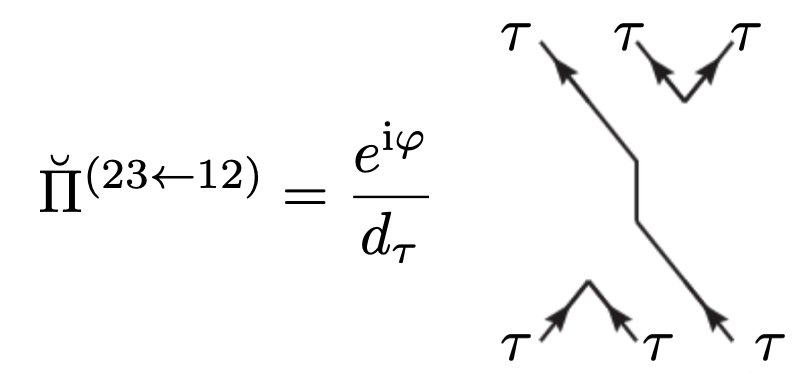}
\end{center}
Next, let us consider promoting the quantum teleportation into braiding. This can be done through the forced measurement using two auxiliary anyons, which is diagrammatically represented as
\begin{center}
    \includegraphics[scale=0.35]{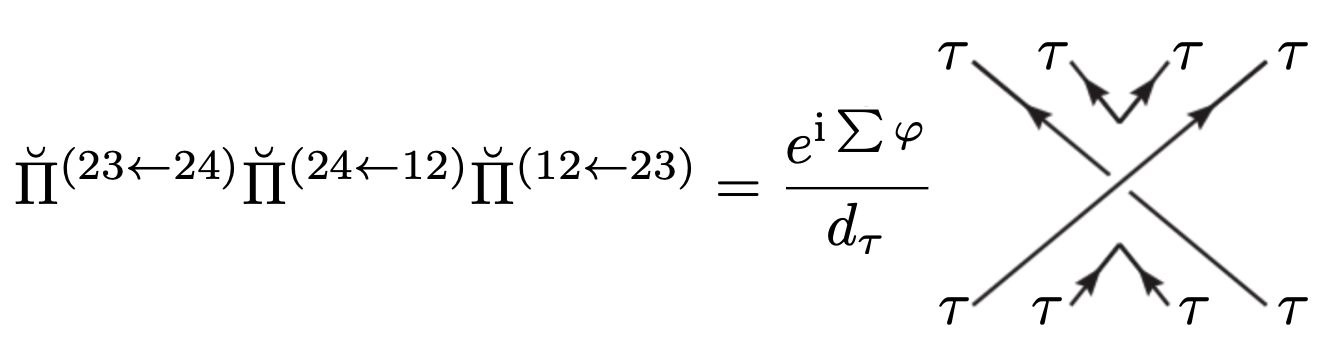}
\end{center}
Consequently, we can use a total number of four anyons to realize a binary braiding.

\bibliography{main.bib}

\begin{thebibliography}{80}
\bibitem [{1()}]{1}%
  \BibitemOpen
  \href@noop {} {}\bibinfo {note} {We note that $\Phi _1$ and $\Phi _2$ are
  primary fields, while $\Gamma $ is the Fibonacci anyon. They share the same
  fusion rules but they are not identified with each other.}\BibitemShut
  {Stop}%
\end{thebibliography}

\end{document}